\documentclass[aps,prb,twocolumn,superscriptaddress]{revtex4-1}

\usepackage{physics}
\usepackage{bbold}
\usepackage{lgrind}        
\usepackage{natbib}    
\usepackage{color}         
\usepackage[pdftex]{graphicx}      
\usepackage{longtable}     
\usepackage{epsf}          
\usepackage{bm}            
\usepackage{amsmath}
\usepackage{thumbpdf}
\usepackage{esint}
\usepackage{subfigure}
\usepackage[colorlinks=true]{hyperref} 
\usepackage{placeins}
\begin{document}

\title{Topological Semimetal Features in the Multiferroic Hexagonal Manganites}


\author{Sophie F. Weber}
\affiliation{Department of Physics, University of California, Berkeley, CA 94720, USA}
\affiliation{Molecular Foundry, Lawrence Berkeley National Laboratory, Berkeley, CA 94720, USA}
\author{Sin\'ead M. Griffin}
\affiliation{Department of Physics, University of California, Berkeley, CA 94720, USA}
\affiliation{Molecular Foundry, Lawrence Berkeley National Laboratory, Berkeley, CA 94720, USA}
\affiliation{Materials Science Division, Lawrence Berkeley National Laboratory, Berkeley, Ca 94720, USA}
\author{Jeffrey B. Neaton}
\affiliation{Department of Physics, University of California, Berkeley, CA 94720, USA}
\affiliation{Molecular Foundry, Lawrence Berkeley National Laboratory, Berkeley, CA 94720, USA}
\affiliation{Kavli Energy NanoScience Institute at Berkeley, Berkeley, CA 94720, USA}

\date{\today}
\begin{abstract}
Using first-principles calculations we examine the band structures of ferromagnetic hexagonal manganites $\mathrm{YXO_3}$ (X=V, Cr, Mn, Fe and Co) in the nonpolar nonsymmorphic $P6_3/mmc$ space group. For $\mathrm{YVO_3}$ and $\mathrm{YCrO_3}$ we find a band inversion near the Fermi energy that generates a nodal ring in the $k_z=0$ mirror plane. We perform a more detailed analysis for these compounds and predict the existence of the topological ``drumhead" surface states. Finally, we briefly discuss the low-symmetry polar phases (space group $P6_3cm$) of these systems, and show they can undergo a $P6_3/mmc \rightarrow P6_3cm$ transition by condensation of soft $K_3$ and $\Gamma_2^-$ phonons. Based on our findings, stabilizing these compounds in the hexagonal phase could offer a promising platform for studying the interplay of topology and multiferroicity, and the coexistence of real-space and reciprocal-space topological protection in the same phase.
\end{abstract}


\maketitle

\section{Introduction}
Ever since their discovery in 1963\cite{BertautE.F.ForratF.Fang1963a}, the hexagonal manganites ($R\mathrm{MnO_3}$, $R=\mathrm{Sc},\mathrm{Y}, \mathrm{In}, \mathrm{Dy}-\mathrm{Lu}$) have attracted great interest by virtue of their combined magnetic and ferroelectric order. The hexagonal manganites undergo an improper ferroelectric transition from a centrosymmetric $P6_3/mmc$ [194] phase ($P\bar{3}c1$ for $\mathrm{InMnO_3}$\cite{Huang2013a}) to the polar $P6_3cm$ [185] structure at around 1000K; they develop a noncollinear antiferromagnetic ground state at much lower temperatures (for the prototypical example of $\mathrm{YMnO_3}$, magnetic ordering sets in around  ~80K\cite{Huang1997}). Multiferroic materials such as the hexagonal manganites are promising for both basic research and for technology due to the possibility for controlling multiple order parameters (via, for example, temperature, magnetic field, or strain) within a single material\cite{Umeda2005a}.\\
\indent Another class of systems of current interest are topological materials, which include, more recently, topological semimetals (TSMs)\cite{Weng2016,Murakami2007}. TSMs exhibit band crossings protected by crystalline and other symmetries. The nodes in TSMs can be either zero-dimensional, as in the case of Dirac and Weyl semimetals\cite{Xu2015,LV2015,Wan2011,Weng2015a,Liu2014a,Liu2014,Wang2013}, or they can form a closed one-dimensional ring, which occurs for nodal line (NL) semimetals\cite{Hu2016,Bian2016,Bian2016b,Neupane2016,Yu2015}. As a consequence of their nontrivial topological character, these three broad categories of TSMs host a wide variety of exotic phenomena including ultrahigh mobility, the chiral anomaly, giant magnetoresistance, and unusual surface states, such as Fermi arcs in Weyl semimetals and two-dimensional `drumhead' states in NLs\cite{Liang2014,Parameswaran2014}.\\
\indent The remarkable properties of TSMs and multiferroic materials have sparked interest in compounds that combine the two properties, i.e. multiferroic systems that are also TSMs in either their high-symmetry nonpolar or low-symmetry polar phases\cite{Tominaga2014,Li2016,Yu2018}. Such compounds can potentially be switched between topological and trivial electronic structure by application of an external field or by tuning temperature through the ferroelectric transition, and they also provide an excellent platform for studying the interplay between the topology, ferroelectricity, and magnetism.\\
\indent There are several arguments for investigating the hexagonal manganite structure as a possible platform for combining multiferroic and TSM properties. First, the synthesis of hexagonal manganites is well-developed both in bulk and in ultrathin epitaxial film form. For example, $R\mathrm{MnO_3}$-type compounds that have an orthorhombic ground state have been grown in the metastable hexagonal structure, primarily via epitaxial stabilization on a hexagonal lattice\cite{Graboy2003}. Second, hexagonal manganites are known to exhibit real-space topological defects in their ferroelectric $P6_3cm$ state, which manifest as adjacent domains of opposite polarization directions, with the vortex phase remaining in the nonpolar $P6_3/mmc$ space group at low temperature\cite{Huang2014}. Such nontrivial real-space topology existing concomitantly with reciprocal-space topological order, i.e the TSM phase, would provide an unprecedented opportunity to explore the interaction between such types of topology. \\
\indent While the prototypical hexagonal manganite $\mathrm{YMnO_3}$ is insulating in both polar and nonpolar phases with its ground state antiferromagnetic (AFM) order, the band structure can be significantly altered by stabilizing ferromagnetic (FM) order. FM order can be achieved, for example, by application of a magnetic field, or by substituting other transition metal (TM) ions for $\mathrm{Mn}^{3+}$ to alter the balance in the competition between in-plane noncollinear AFM order and slight out-of-plane canting which is common in the hexagonal manganites\cite{Solovyev2012}.\\
\indent In this work, we undertake a first-principles study of the electronic band structure and its topology in compounds isostructural to $\mathrm{YMnO_3}$ with FM ordering. Specifically, in addition to $\mathrm{YMnO_3}$, we investigate four other compounds in which the $\mathrm{Mn}^{3+}$ cation has been substituted with $\mathrm{V}^{3+}$, $\mathrm{Cr}^{3+}$, $\mathrm{Fe}^{3+}$ and $\mathrm{Co}^{3+}$ in order to shift the Fermi level systematically. We predict that nonpolar hexagonal manganites $\mathrm{YVO_3}$ and $\mathrm{YCrO_3}$ have band crossings very close to the Fermi level, and in fact feature topological nodal lines in the $k_z=0$ plane that are protected by a mirror symmetry. We also predict that they should undergo a ferroelectric (FE) $P6_3/mmc\rightarrow P6_3cm$ transition characteristic of the traditional $R\mathrm{MnO_3}$ compounds. Stabilizing this set of compounds in the hexagonal structure should hence provide new opportunities for studying the interaction between topological and multiferroic order. 

\section{Results}

\subsection{\label{method}Methodology}
For our first-principles density functional theory (DFT) calculations, we employ the Vienna \emph{ab intitio} simulation package (VASP)\cite{Kresse1996} with generalized gradient approximation (GGA) using the Perdew-Burke-Ernzerhof (PBE) functional\cite{Perdew1996} and projector augmented-wave (PAW) method\cite{Blochl1994}. We treat $4s, 4p, 5s$ and $4d$, and $2s$ and $2p$ electrons explicitly as valence for $\mathrm{Y}$ and $\mathrm{O}$, respectively. For the five transition metals $\mathrm{V}$-$\mathrm{Co}$, we include $3p$ as well as the valence $d$ and $s$ electrons. To account for the localized nature of the $d$ electrons in the transition metal cations, we add a Hubbard U correction (GGA+U)\cite{Perdew1986}. We apply the rotationally invariant version of GGA+U by Dudarev et al.\cite{Dudarev1998}, and for ease of comparison we choose a $U$ of 3 eV for all elements, a value consistent with previous literature\cite{Lany2013} (see supplementary material for further discussion of our GGA+U calculations). We use an energy cutoff of 800 eV for our plane wave basis set, with a Gamma-centered $\mathbf{k}$-point mesh of $16\times16\times6$ for the 10-atom nonpolar unit cell and $8\times8\times6$ for the 30-atom polar unit cell. Starting from the structures in the Materials Project database\cite{Jain2013}, we relax lattice parameters and internal coordinates for all structures until forces on the atoms are less than $0.001$ eV/\AA. We use collinear spin-polarized calculations to account for the finite magnetic moments of the transition metal (TM) ions. We do not include spin-orbit coupling (SOC) unless explicitly stated. When relevant, we approximate the noncollinear AFM order inherent to the hexagonal manganites\cite{Filippetti2001} with a collinear G-type AFM ordering (GAFM), consisting of a two up, one down (one up, two down) pattern in a given 30-atom supercell for the upper (lower) basal plane\cite{Zhong2009} (see Figure \ref{fig:GAFM}). Finally, for all topological analysis we use a tight-binding model constructed from our DFT-GGA+U calculations using maximally localized Wannier functions (MLWFs)\cite{Marzari1997,Mostofi2014} as our basis states. The tight-binding model is then used as input in the WannierTools package\cite{Wu2017} to calculate surface states as well as presence and location of the nodal rings.

\subsection{\label{structure}Nonpolar $P6_3/mmc$ Crystal Structure and Energetics}
\begin{figure*}
\subfigure[]{\label{fig:NPstruct}}\includegraphics[width=.98 \columnwidth]{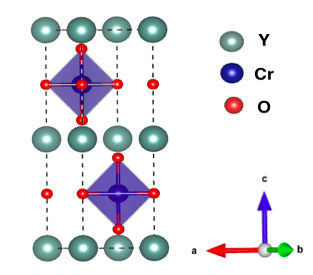}
\subfigure[]{\label{fig:GAFM}}\includegraphics[width=.98 \columnwidth]{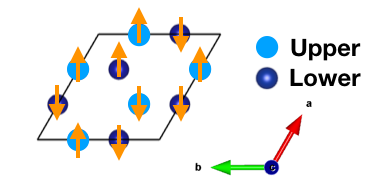}
\caption{(a) Primitive cell for nonpolar  hexagonal $\mathrm{YCrO_3}$, with centrosymmetric space group $P6_3/mmc$ [194]. The primitive cell consists of two Y atoms (green), two Cr atoms (blue), and six O atoms (red) (the structures of all other compounds studied in this paper are qualitatively identical). (b) Depiction of the GAFM in-plane magnetic ordering which we use as a collinear approximation to the true noncollinear antiferromagnetism in the hexagonal manganites. Note that the 10-atom $P6_3/mmc$ primitive cell is tripled to accommodate this ordering.}
\end{figure*}

To begin, we focus on the centrosymmetric, nonpolar crystal structure of the hexagonal manganites in the $P6_3/mmc$ space group. This $\mathrm{YXO_3}$ ($\mathrm{X}=\mathrm{V},\mathrm{Cr}, \mathrm{Mn}, \mathrm{Fe}$ and $\mathrm{Co}$) structure is shown in Figure \ref{fig:NPstruct} ($\mathrm{YCrO_3}$ is chosen as the example). The primitive cell consists of two TM atoms, two yttrium atoms, and six oxygen atoms. The TM ions $\mathrm{X^{3+}}$ are five-fold coordinated by oxygen $\mathrm{O^{2-}}$, forming trigonal bipyramids, and they lie in the $z=\frac{1}{4}$ and $z=\frac{3}{4}$ planes. The yttrium $\mathrm{Y^{3+}}$ ions are sandwiched between, in the $z=0$ and $z=\frac{1}{2}$ planes.\\
\indent While all of our subsequent calculations are performed assuming FM order, we also perform relaxations using the GAFM configuration in order to examine the relative energies of the two magnetic states. We note that at the high temperatures for which the nonpolar $P6_3/mmc$ space group is naturally favored over the polar $P6_3cm$ space group, the structure is paramagnetic. However, it may be possible to stabilize the nonpolar structure at low temperatures, for example by alloying or introducing defects\cite{ Griffin2017}. The results of our calculations for both orderings are given in Table \ref{tab: params}. To date, the only crystal in Table \ref{tab: params} which has been synthesized in bulk $P6_3/mmc$ structure is $\mathrm{YMnO_3}$, with reported lattice parameters $a=3.61$ \AA{} and $c=11.39$ \AA\cite{Lukaszewicz1974}. Comparing this to our relaxed FM GGA+U results of $a=3.617$ \AA{} and $c=11.366$ \AA{} suggests that our GGA+U calculations will be good predictors of experimental lattice constants of the other four compounds.\\
\indent We also report energy differences between FM and GAFM orderings in Table \ref{tab: params}. These values may be viewed as a guide since the frustrated collinear GAFM order is an approximation to that of noncollinear AFM. Nonetheless, comparison with collinear GAFM should be useful for predicting the relative ease of stabilizing the FM state in these compounds, for example by application of a magnetic or electric field\cite{Lottermoser2004}. Specifically, FM ordering becomes more stable relative to GAFM the further to the left we move on the periodic table, so achieving FM order should be most feasible in the $\mathrm{V}^{3+}$ and $\mathrm{Cr}^{3+}$ compounds. Note that when GAFM is enforced, for $\mathrm{X}=\mathrm{Cr}$-$\mathrm{Co}$ the relaxed $\mathrm{O}$-$\mathrm{X}$-$\mathrm{O}$ bond angle between apical and in-plane oxygen atoms differs by less than $0.5^{\circ}$ from the ideal $90^{\circ}$. However, in the case of $\mathrm{YVO_3}$ the enforced magnetic frustration results in non-uniform magnetic moments on the inequivalent $\mathrm{V}$ atoms, leading to a large and unrealistic distortion of the bond angles by as much as $15^{\circ}$. Including SOC and allowing $\mathrm{YVO_3}$ to relax to the noncollinear AFM should remove this artifact, but to be consistent with the other compounds in Table \ref{tab: params} we include our results for the collinear GAFM structure; the parameters and energetics for $\mathrm{YVO_3}$ with this enforced magnetic order relative to the other four compounds should be interpreted with appropriate caution.


 \begin{table}
 \caption{\label{tab: params}Lattice constants (for the primitive 10-atom unit cell), energy per formula unit (f.u) for FM and collinear GAFM ordering, and $\Delta E=E_{FM}-E_{GAFM}$ for $\mathrm{YXO_3}$ in the $P6_3/mmc$ space group after full optimization with GGA+U. *As mentioned in the main text, the inherent frustration of the GAFM ordering on a triangular lattice has a strong effect on the bond angles of $\mathrm{YVO_3}$. We include the relaxed GAFM result for completeness but with the caveat that the distortion may be unphysical.}
 \begin{ruledtabular}
 \begin{tabular}{| c | c | c | c | c | c |}
 \hline
 & $\mathrm{YVO_3}$ & $\mathrm{YCrO_3}$ & $\mathrm{YMnO_3}$ & $\mathrm{YFeO_3}$ & $\mathrm{YCoO_3}$ \\ \hline
FM & & & & & \\
$a$ (\AA) & 3.496 & $3.510$ & $3.617$ & $3.566$ & $3.640$ \\
$c$ (\AA)& $12.382$ & $12.041$ & $11.366$ & $11.762$ & $11.193$ \\
E/f.u (eV) & $-42.366$ & $-42.242$ & $-42.188$ & $-40.223$ & $-37.218$\\ \hline
GAFM & & & & &\\
$a$ (\AA) & 3.561* & $3.525$ & $3.609$ & $3.548$ & $3.608$ \\
$c$ (\AA)& $11.912*$ & $12.010$ & $11.359$ & $11.798$ & $11.272$ \\
E/f.u (eV) & $-42.341$ & $-42.062$ & $-42.182$ & $-40.407$ & $-37.512$\\ \hline
$\Delta E$ (eV) & $-0.025$  & $-0.180$ & $-0.006$ & $+0.184$ & $+0.294$ \\
\hline
\end{tabular}
\end{ruledtabular}
\end{table}

\subsection{\label{band_structs} Semimetal Features in Ferromagnetic Band Structures}
\begin{figure*}
\subfigure[]{\label{fig:V}}\includegraphics{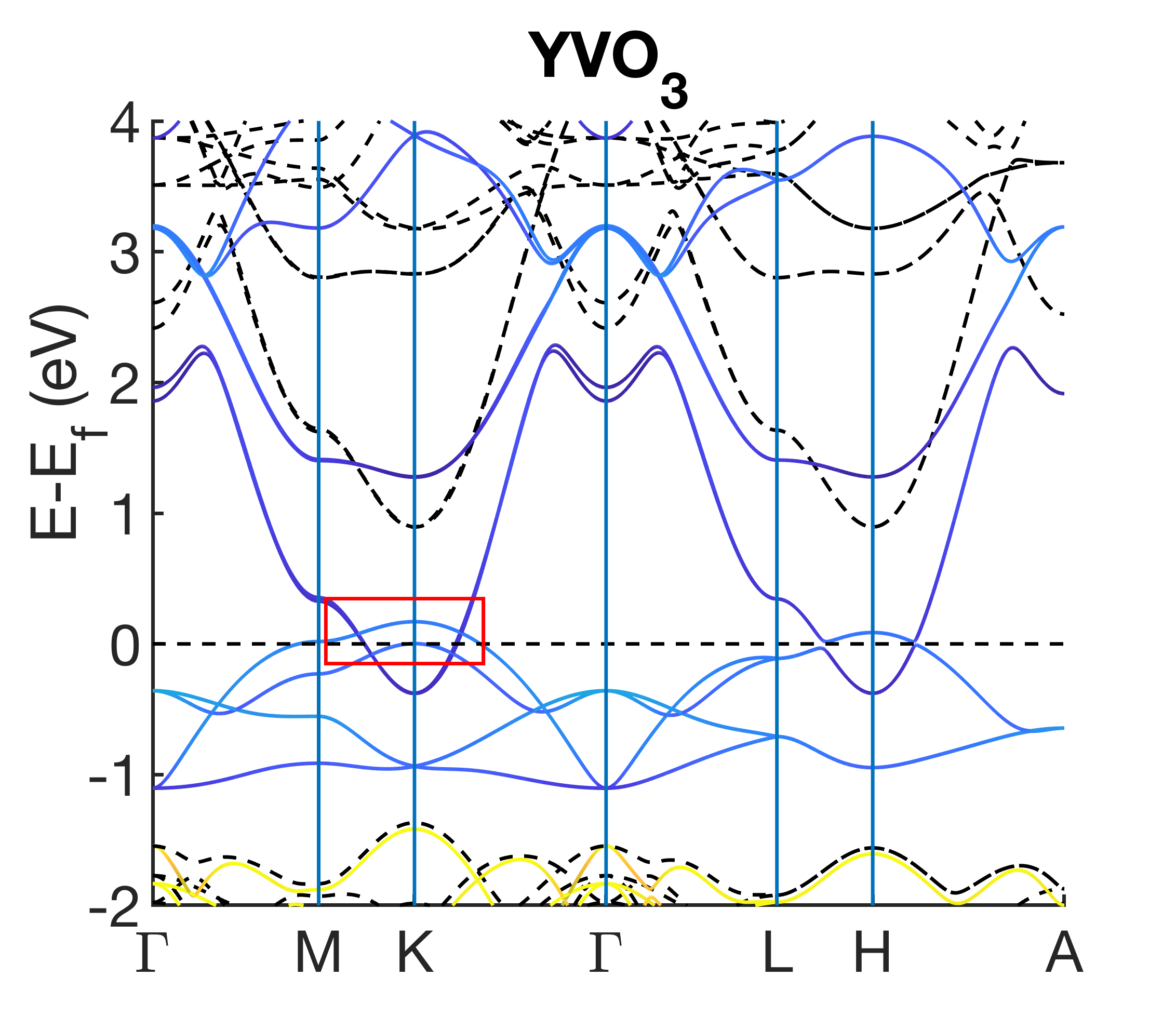}
\subfigure[]{\label{fig:Cr}}\includegraphics{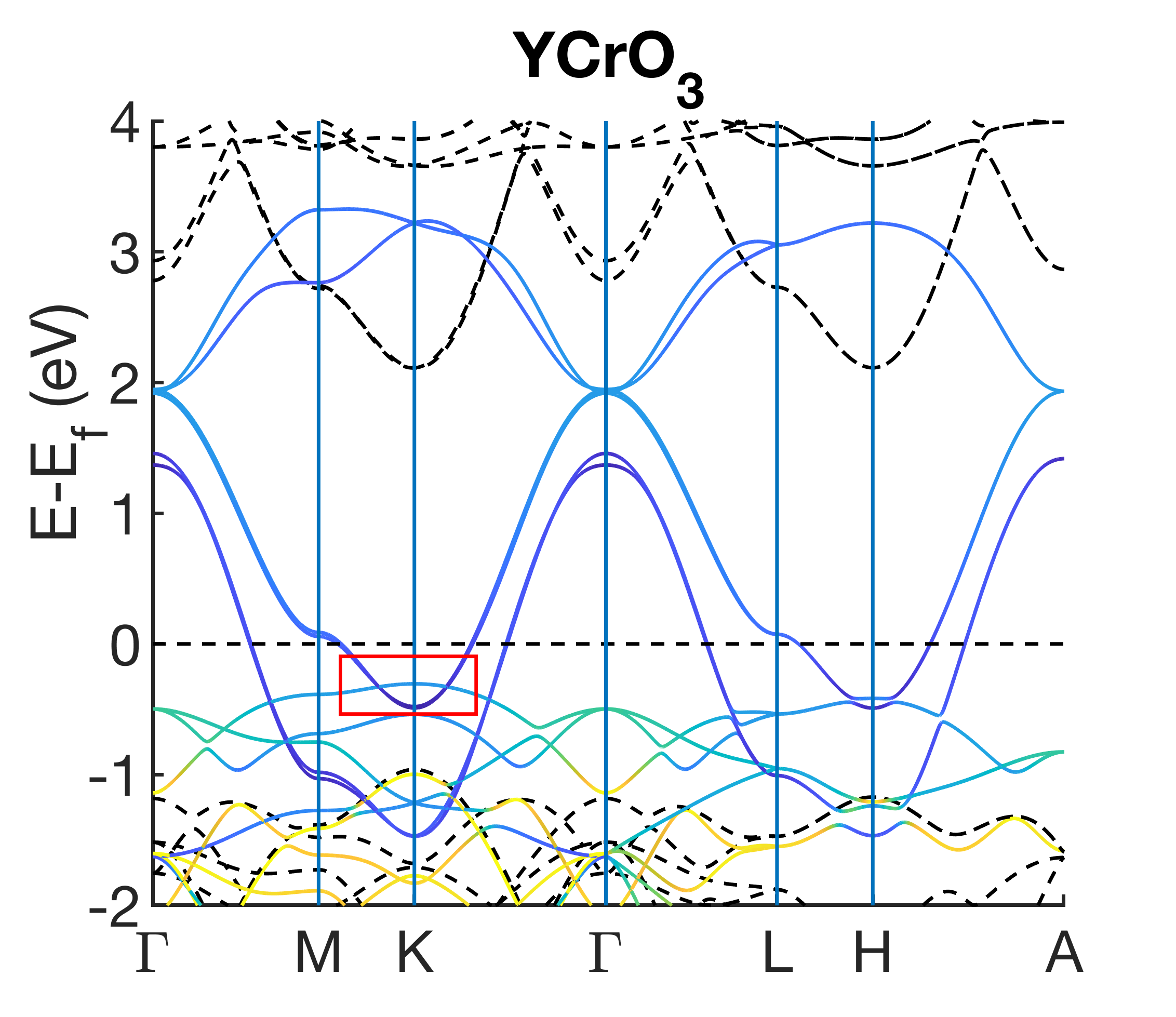}
\subfigure[]{\label{fig:Mn}}\includegraphics{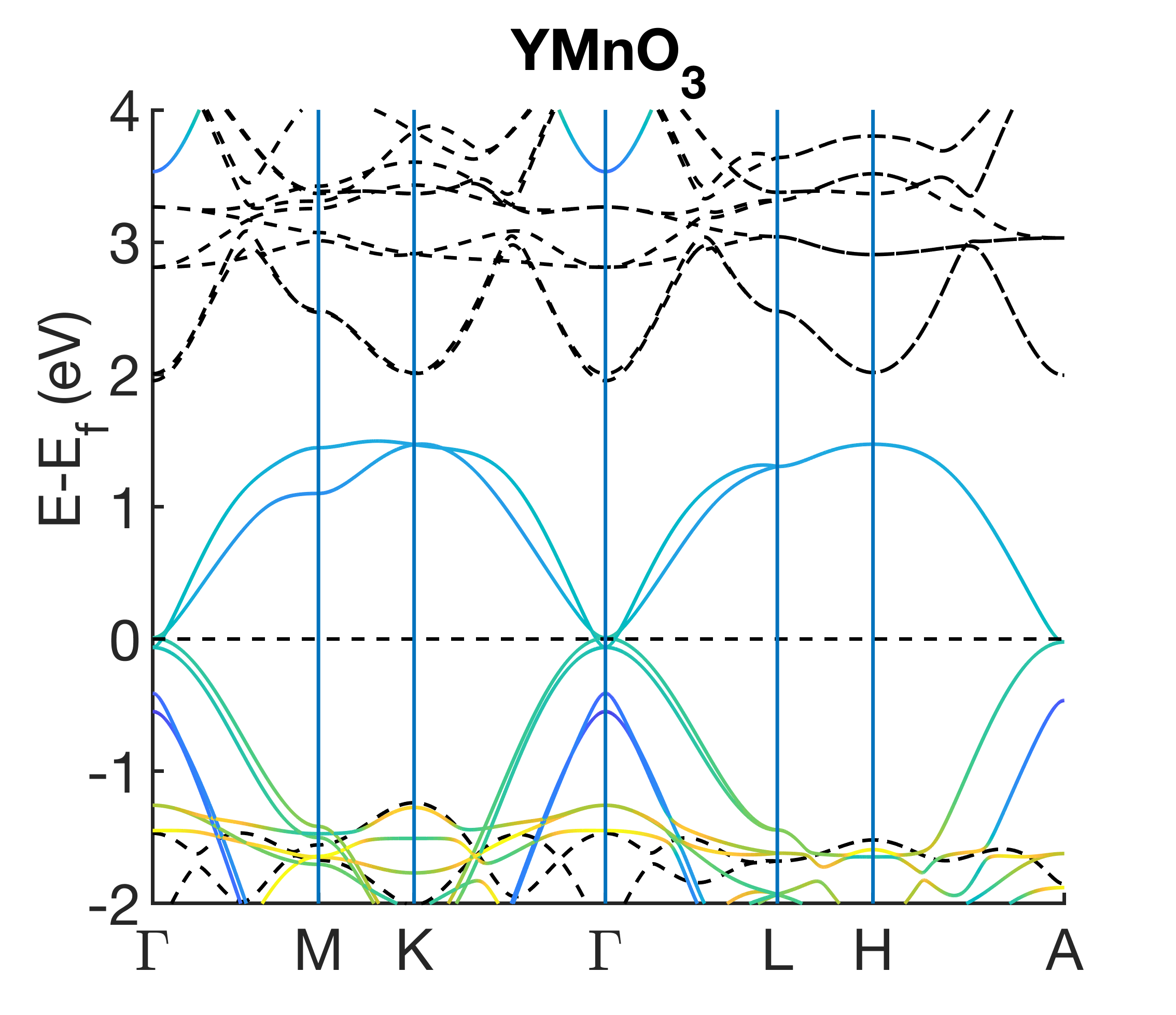}
\subfigure[]{\label{fig:Fe}}\includegraphics{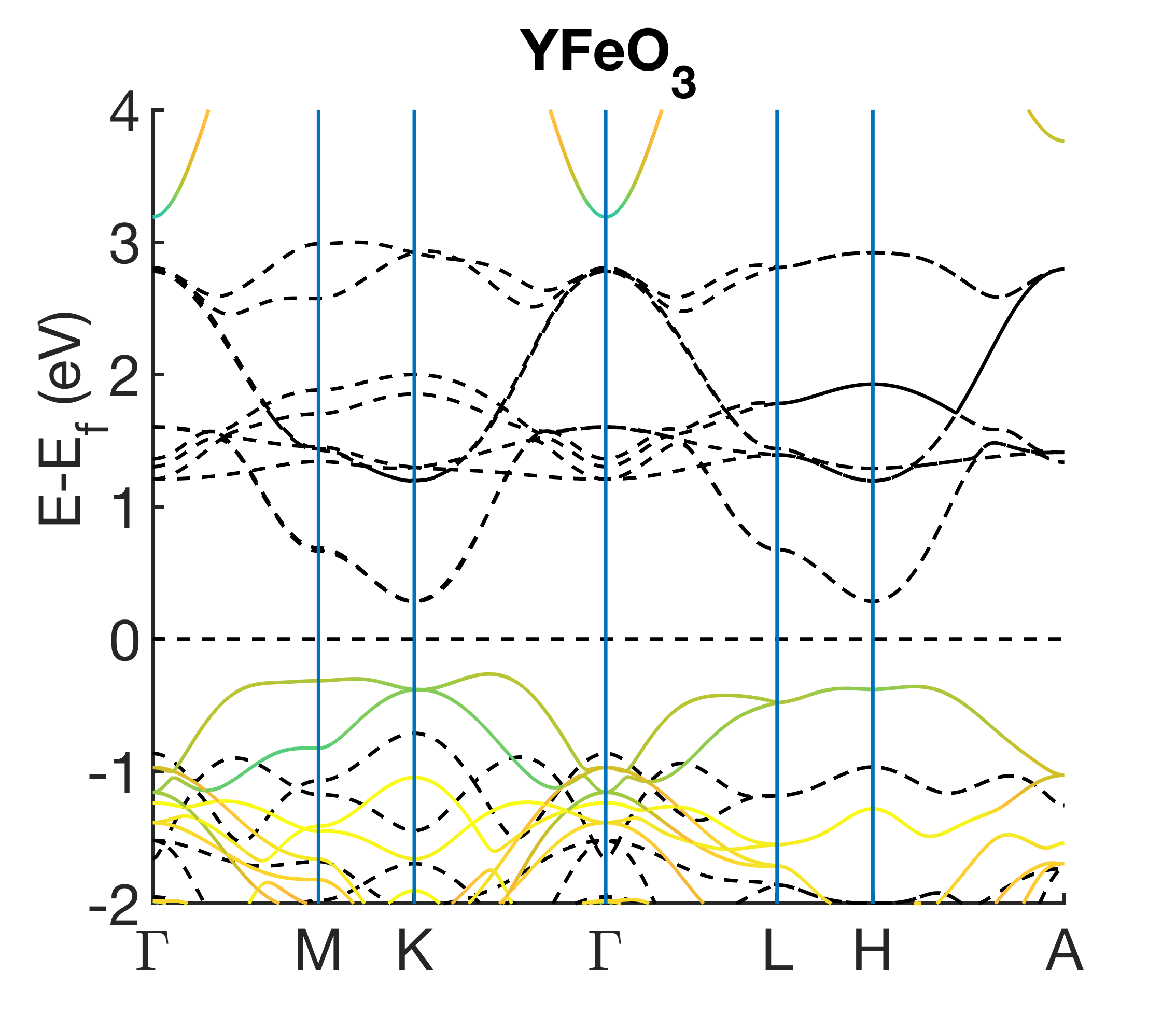}
\subfigure[]{\label{fig:Co}}\includegraphics{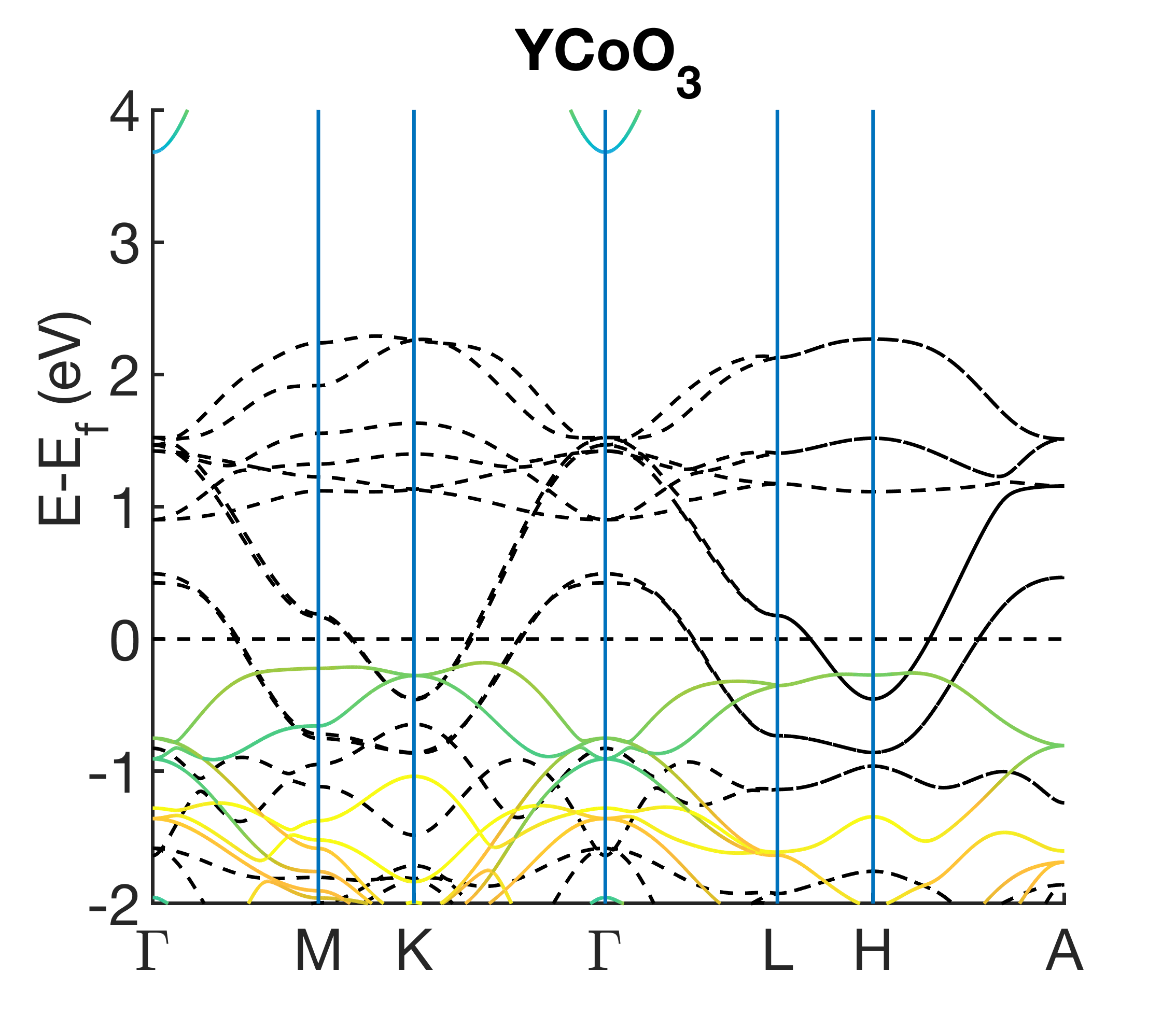}
\subfigure{\label{fig:bar}}\includegraphics{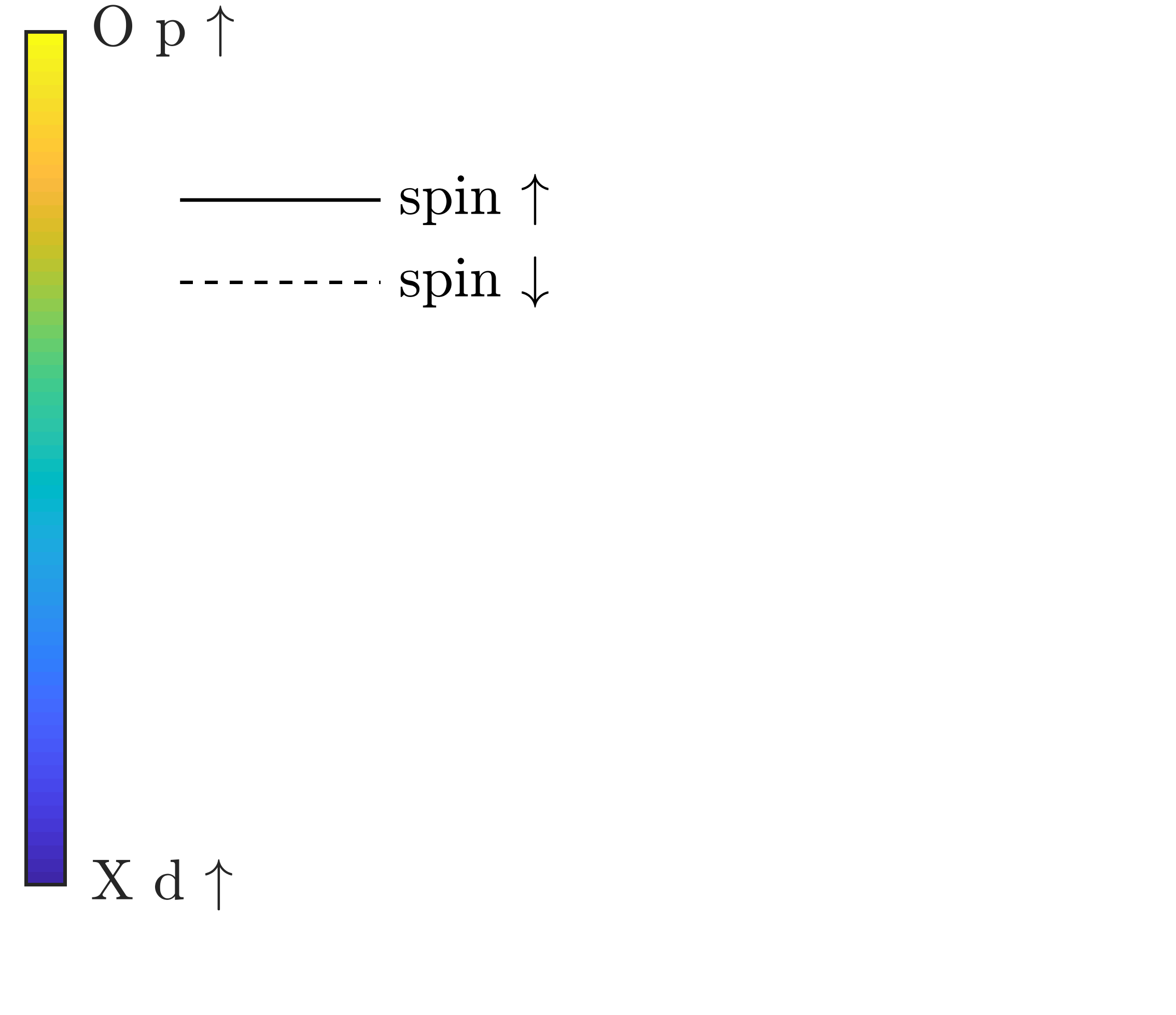}
\caption{\label{fig: orbplots} Orbital-projected DFT-GGA+U ($U=3$ eV) band structures for spin up bands in the ferromagnetic $P6_3/mmc$ $\mathrm{YXO_3}$ compounds ($\mathrm{X}=\mathrm{V}-\mathrm{Co}$), with spin down bands included without projections. The bands near the Fermi energy are composed of $\mathrm{X}$ $d$ states (where $\mathrm{X}$ is the relevant transition metal ion) and $\mathrm{O}$ $p$ states, with negligible $\mathrm{Y}$ character. Color scale varies from dark blue for purely $\mathrm{X}$ $d$ character to yellow for purely $\mathrm{O}$ $p$ character. The Fermi level is marked by the dashed black line. Panels (a)-(e) correspond to $\mathrm{YVO_3}$, $\mathrm{YCrO_3}$, $\mathrm{YMnO_3}$, $\mathrm{YFeO_3}$, and $\mathrm{YCoO_3}$, respectively.}
\end{figure*}

\begin{figure}
\subfigure[]{\label{fig:YVO3_zoom}}\includegraphics{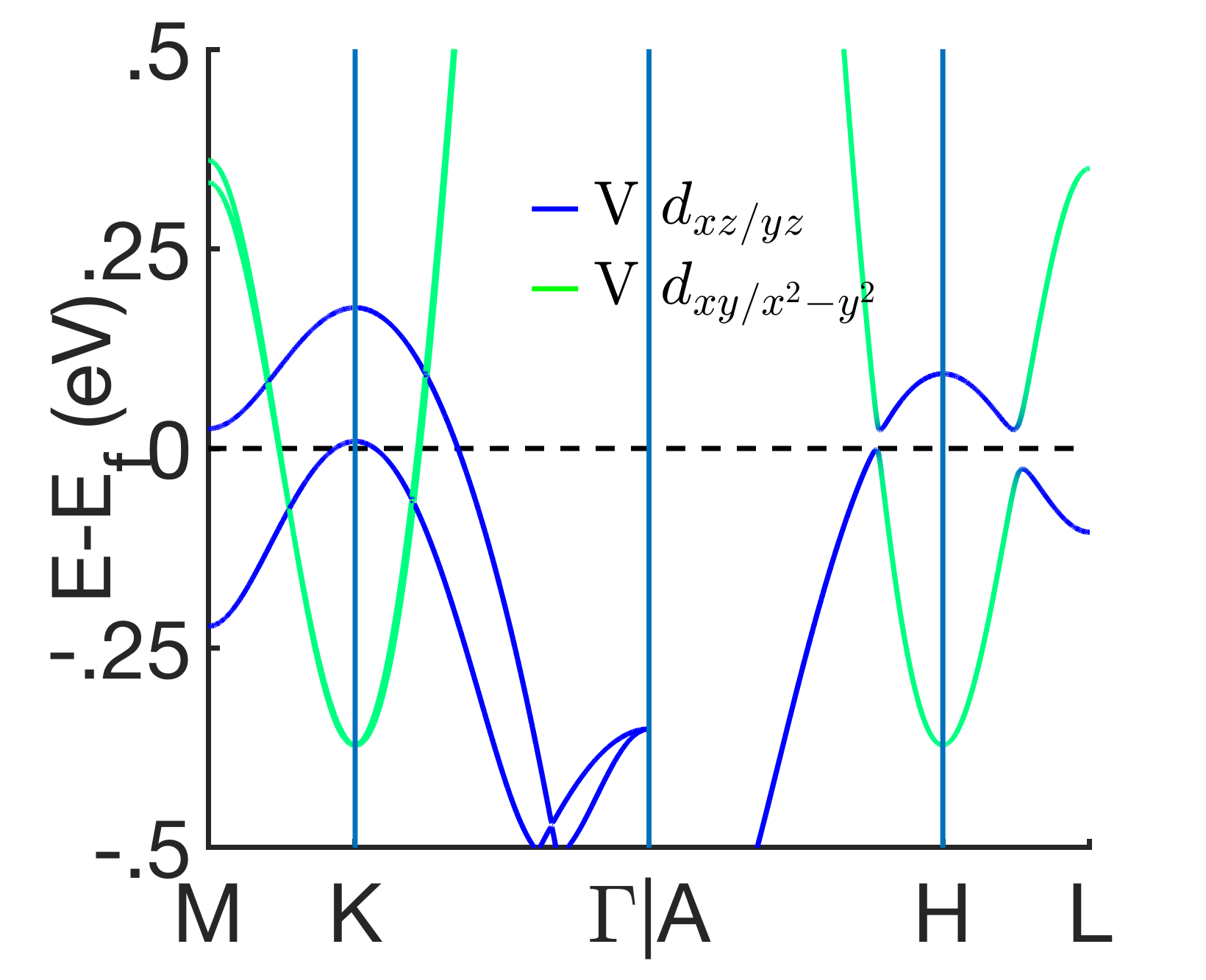}
\subfigure[]{\label{fig:YCrO3_zoom}}\includegraphics{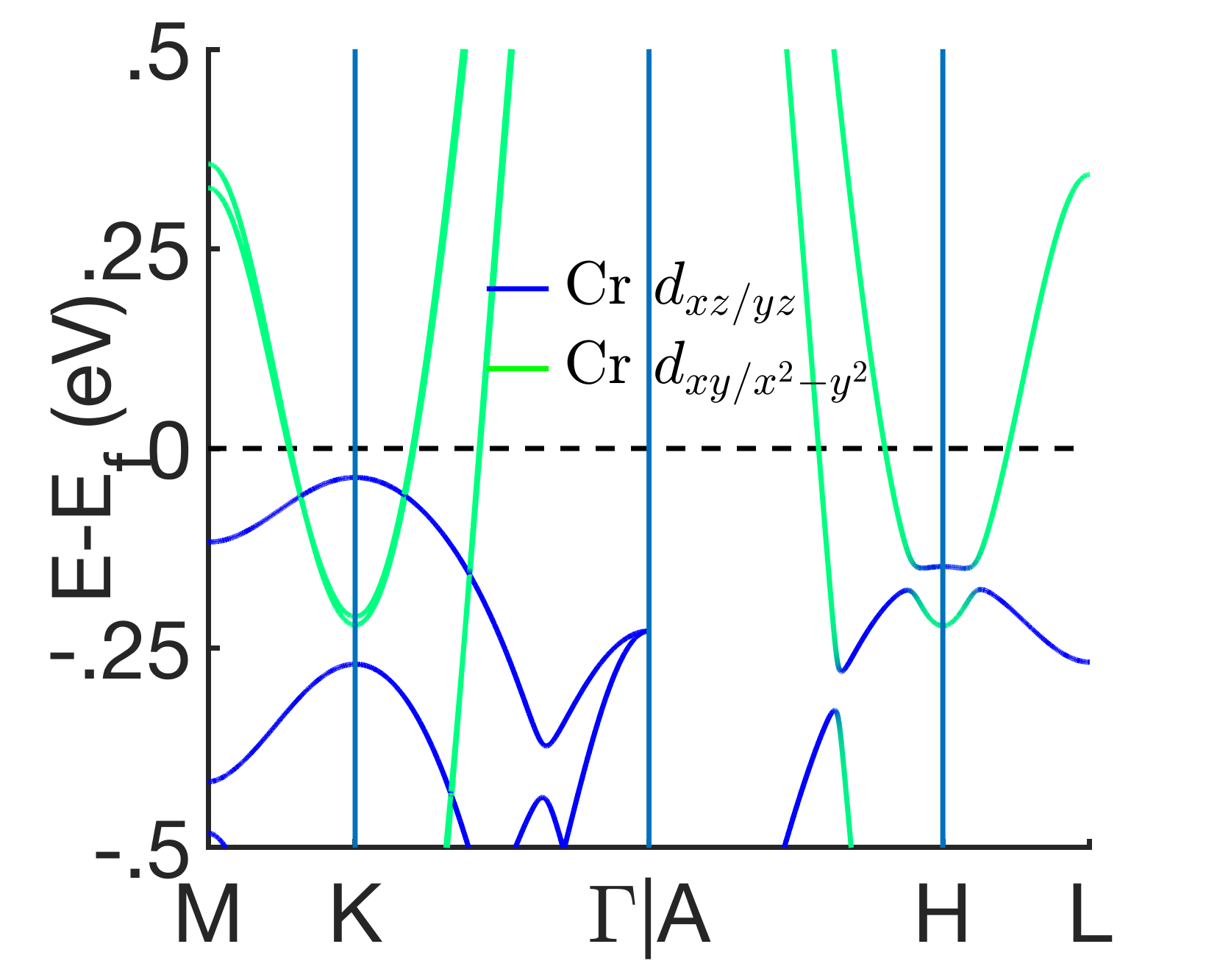}
\subfigure[]{\label{fig:BZ}}\includegraphics[width=\columnwidth]{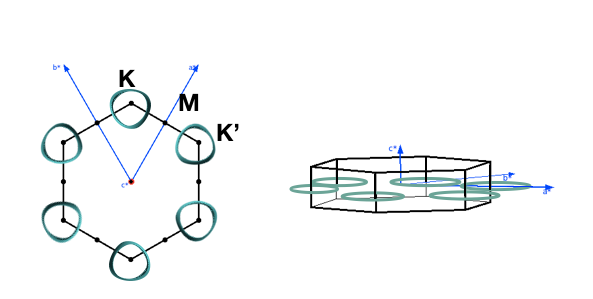}
\caption{\label{fig: nodes} Analysis of band crossings in Figure \ref{fig: orbplots}. (a) and (b) show zoomed-in band structures of $\mathrm{YVO_3}$ and $\mathrm{YCrO_3}$ respectively about the $K$ and $H$ points, with the orbital character decomposed into planar ($d_{xy}/d_{x^2-y^2}$) and z-oriented ($d_{xz}/d_{yz}$) d states. (c) Isoenergy contours (specifically for $E=-0.38$ eV for $\mathrm{YCrO_3}$) in the hexagonal 3D Brillouin zone. $\mathrm{YVO_3}$ is qualitatively identical.}
\end{figure}

In Figure \ref{fig: orbplots} we present the GGA+U band structures for the $P6_3/mmc$ $\mathrm{YXO_3}$ compounds in the FM configuration in the absence of SOC (see supplementary material for GAFM band structures). Because they dominate the states near the Fermi level, we focus on the spin up bands and plot their orbital-projected character. (The spin down bands are included with dashed black lines and without orbital projection.) The spin up bands near the Fermi level are composed almost exclusively of transition metal $\mathrm{X}$ $d$ states and $\mathrm{O}$ $p$ states. Going from left to right across the $3d$ elements, we observe a simultaneous upwards shift of the Fermi level and a lowering in energy of the $\mathrm{X}$ $d$ states toward the $\mathrm{O}$ $p$ states, leading to greatest hybridization for $\mathrm{YFeO_3}$. In $\mathrm{YVO_3}$ and $\mathrm{YCrO_3}$, the uppermost $d$ states have started to invert energies with the lower states of mixed $d$ and $p$ character; in particular, for both we compute a band inversion resulting in linear Dirac nodes centered at the $K$ point $(\frac{1}{3},\frac{1}{3},0)$, boxed in red (upon further inspection the apparent inversion at $H$ is actually an anticrossing). For $\mathrm{YVO_3}$, the crossings at $K$ are about $80$ meV above Fermi level, whereas for $\mathrm{YCrO_3}$ they are about $300$ meV below.\\
\indent We note that GGA+U Kohn-Sham eivenvalues can only approximate single-particle excitations and band structure. Therefore it is reasonable to question whether for GGA+U, and specifically for $U=3$ eV, our approach to computing the band structure near the Fermi energy, specifically the band inversions responsible for the nodes in $\mathrm{YVO_3}$ and $\mathrm{YCrO_3}$, will be predictive. Based on prior calculations for similar oxide systems with $\mathrm{V}$ and $\mathrm{Cr}$ in the same $3^+$ oxidation state, a $U$ of $3$ eV can lead to band structures that nearly reproduce experimental gaps (see the supplemental material for a detailed discussion). Thus, we have reason to expect our $U=3$ eV calculations will be qualitatively accurate for the band inversions near the Fermi energy. \\
\indent We now further analyze the band structure and topology of $\mathrm{YVO_3}$ and $\mathrm{YCrO_3}$. Orbital decompositions of the two inverted bands reveals that one band is composed of mostly $\mathrm{V}$/$\mathrm{Cr}$ $d_{xz}$ and $d_{yz}$ states and the other of $d_{xy}$ and $d_{x^2-y^2}$ states. Plotting only these projections, it is clear that at $K$ the bands cross with no mixing whereas at the $(\frac{1}{3},\frac{1}{3},\frac{1}{2})$ $H$ point they hybridize, exchange character, and very slightly gap out (see Figures \ref{fig:YVO3_zoom} and \ref{fig:YCrO3_zoom}). In both cases we find that the non-gapped crossings in fact form a pair of nodal lines (NLs) lying in the $k_z=0$ plane, one centered at $K$ and the other at $K'$ (see Figure \ref{fig:BZ}).\\
\indent The mirror plane symmetry $\mathcal{M}_z$ centered at $z=\frac{1}{4}$ is responsible for the protection of the NLs. (Note that with FM ordering and no SOC the magnetic space group is identical to the crystal space group $P6_3/mmc$). The action of $\mathcal{M}_z$ in reciprocal space is

\begin{equation}
\mathcal{M}_z: (k_x,k_y,k_z)\rightarrow(k_x,k_y,-k_z).
\label{eq:recspacemirror}
\end{equation}

\begin{figure}
\subfigure[]{\label{fig:SOC_zax}}\includegraphics{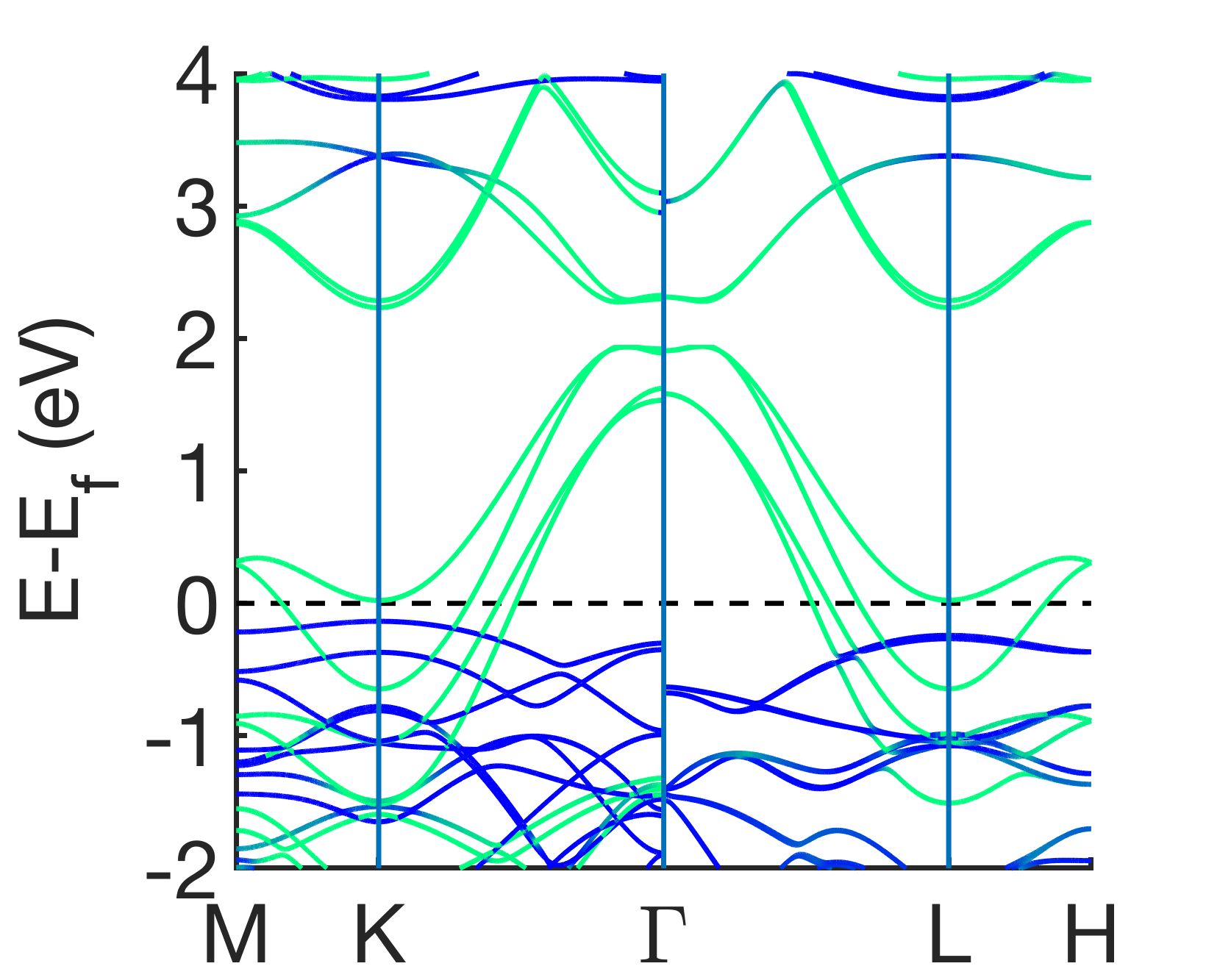}
\subfigure[]{\label{fig:SOC_xax}}\includegraphics{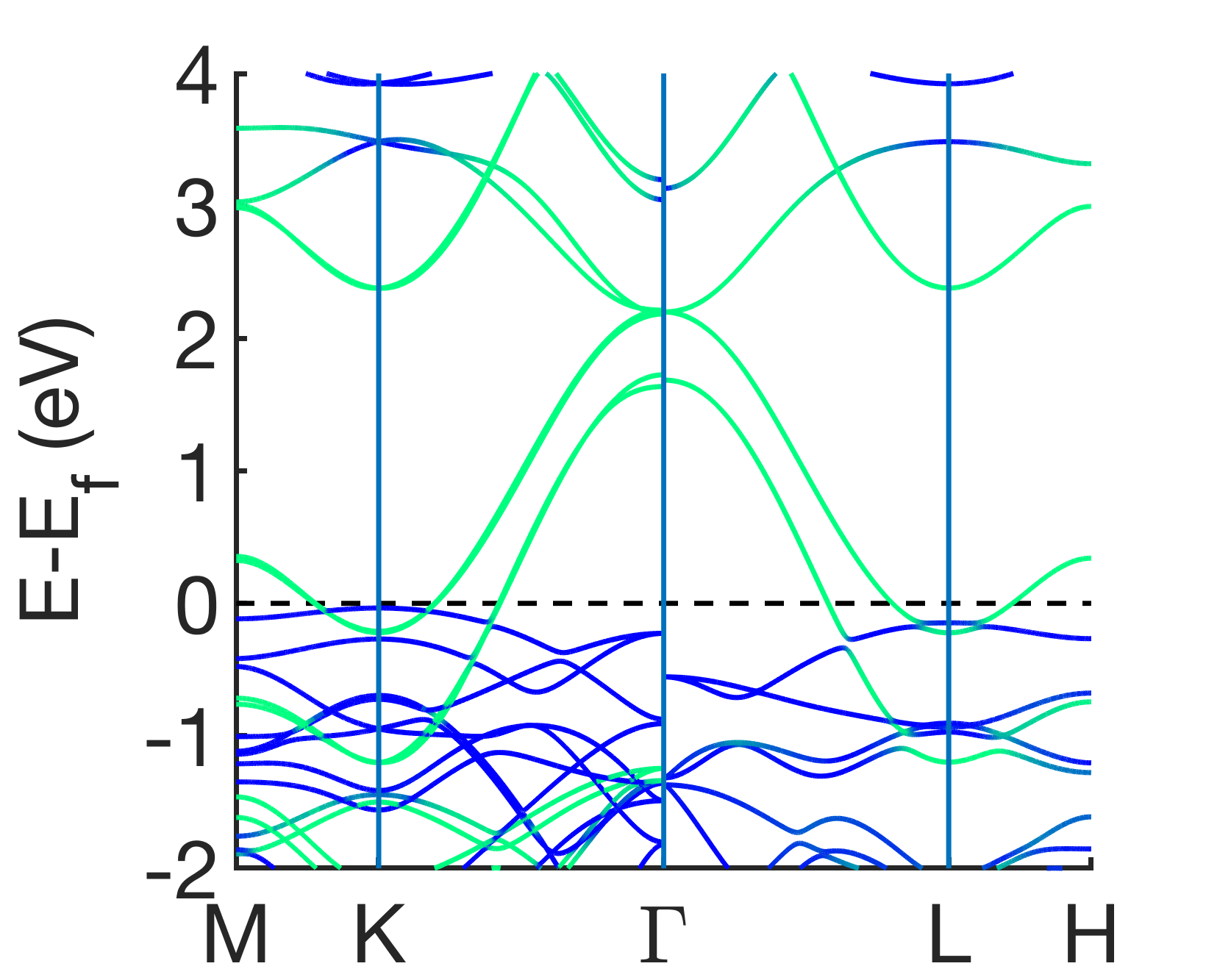}
\subfigure[]{\label{fig:SOC_zax_zoom}}\includegraphics{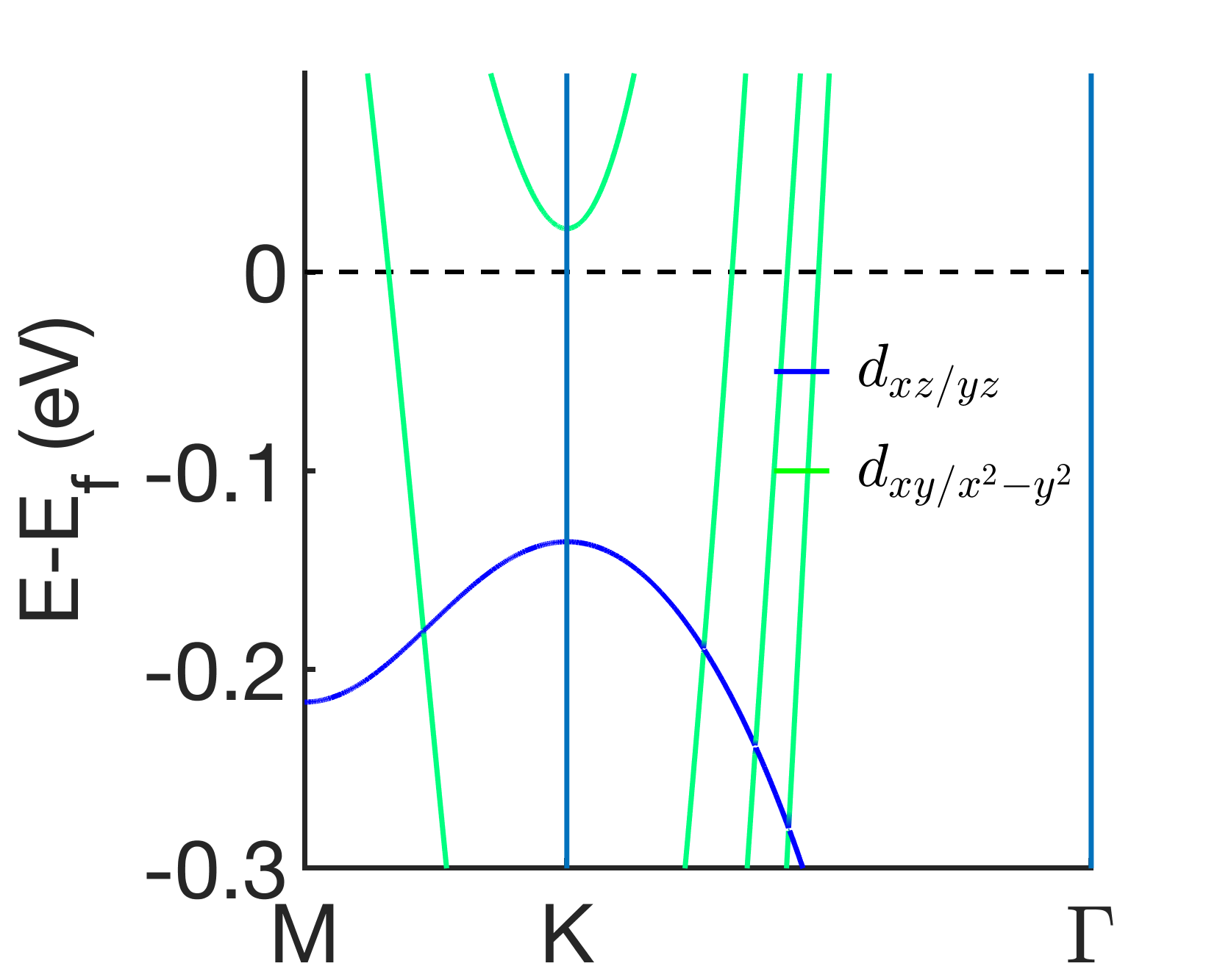}.
\subfigure[]{\label{fig:SOC_xax_zoom}}\includegraphics{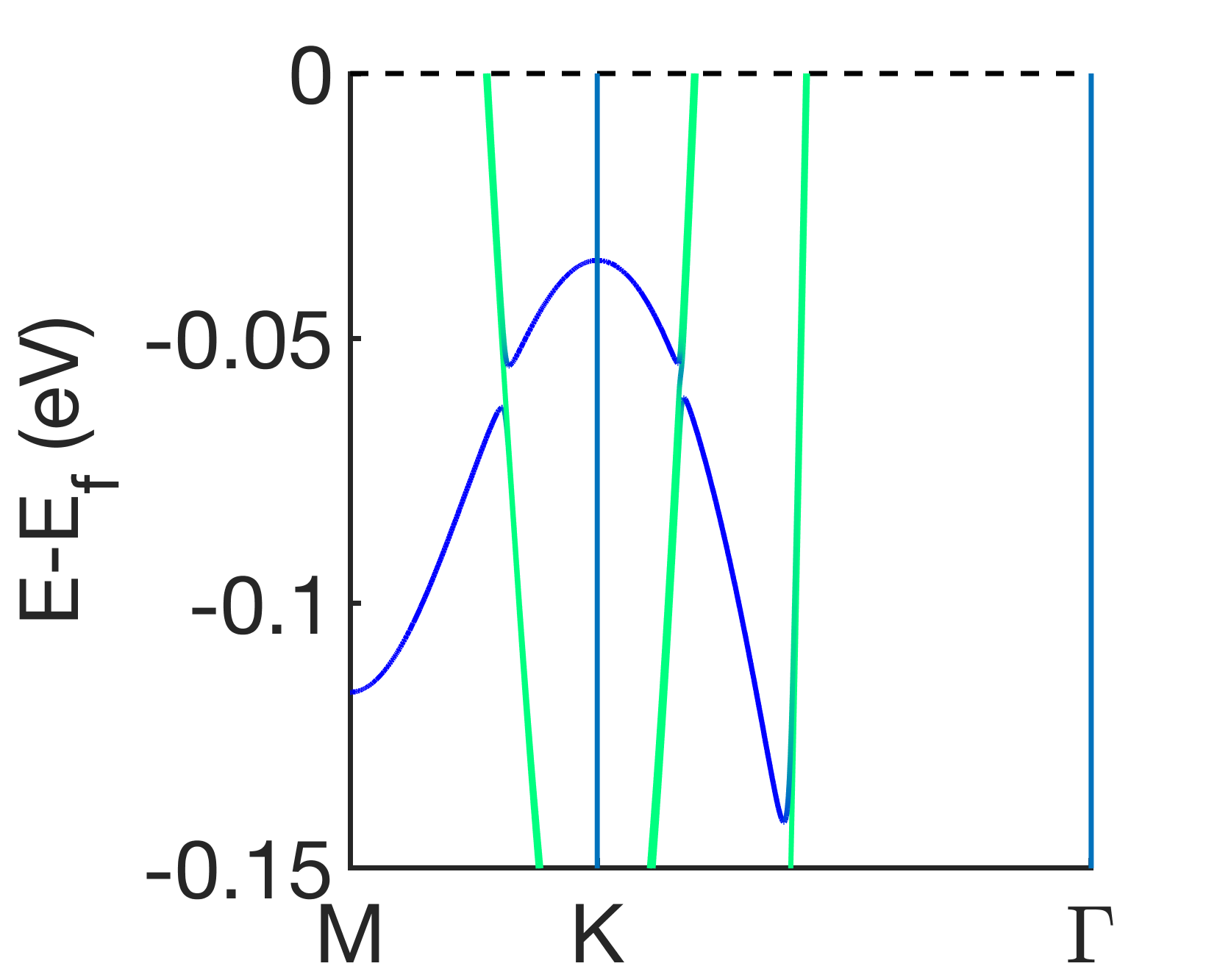}
\caption{\label{fig: SOC}DFT-GGA+U band structures with SOC, using $\mathrm{YCrO_3}$ as the example. (a) and (b) show full band structures with SOC included and spin quantization along the [001] and [100] directions respectively. (c) and (d) show the zoomed-in portions of (a) and (b) around the $K$ where the topological NLs are centered in the collinear spin case. In (d) the NL crossings are still robust with the [001] spin orientation, whereas a very small gap forms between one of the conduction bands and the valence band in (e) with [100] spin orientation.}
\end{figure}

\noindent Thus $k_z=0$ and $k_z=\frac{\pi}{c}$ planes are invariant under $\mathcal{M}_z$ and can be labeled by its eigenvalues, which are $\pm{1}$ in the absence of SOC. If two bands with opposite mirror eigenvalues cross on one of these planes due to a band inversion, their crossing is symmetry-protected and they form a closed loop of Dirac nodes (see supplementary material for proof that the band inversion necessarily results in a one-dimensional NL rather than discrete Dirac points). This is the case for $K$ in the $k_z=0$ plane. However, if the bands have the same eigenvalues they will mix and gap out\cite{Fang2015}, which occurs on the $k_z=\frac{\pi}{c}$ plane where $H$ lies. In the supplementary material we construct an explicit tight-binding model to calculate the $\mathcal{M}_z$ eigenvalues throughout the Brillouin zone and thus verify our observations.\\
\indent Let us now consider what happens when we include spin-orbit coupling (SOC). With SOC, spin and orbital degrees of freedom are coupled and symmetry operators act on both Hilbert spaces simultaneously. Notably for us, a mirror plane symmetry becomes the combination of (a) a reflection of the spatial coordinates about the mirror plane, and (b) a $\pi$ rotation of the spin coordinates about the axis perpendicular to the mirror plane\cite{Weng2015a}. Thus, depending on the spatial orientation of the spins, a mirror plane symmetry may either be broken or preserved when SOC is taken into account. For our nonpolar hexagonal manganites, let us first examine the case where the spin orientation is along the [001] axis. In this example, the magnetic point group symmetry is reduced from $D_{6h}$ in the collinear case to $C_{6h}$. In $C_{6h}$, the total mirror symmetry is still preserved, since rotating the spins $180^o$ about the $z$ axis leaves them invariant; thus, the NL should still be protected in the $k_z=0$ plane in this case. The only difference from the non-SOC case is the functional form of $\mathcal{M}_z$ due to the requirement that the operator must now be antiunitary, such that spin up (down) bands pick up a factor of $+i$ ($-i$) when acted on by $\mathcal{M}_z$ (we include a tight-binding model which incorporates SOC in addition to the collinear tight-binding models in our supplementary material). Taking the example of $\mathrm{YCrO_3}$, we plot the band structure with SOC for [001] oriented spins and as expected the crossings are still robust.\\
\indent If we choose, on the other hand, to orient the spins such that they have a component perpendicular to the [001] axis, say in the [100] direction, the magnetic point group is reduced to $C_{2h}$.  Now the action of the mirror operator will still leave the orbital coordinates in the $k_z=0$ plane unchanged, but it will send a spin with components $(s_x,s_y,s_z)=(1,0,0)$ to $(s_x,s_y,s_z)=(-1,0,0)$. Hence the mirror plane is no longer a symmetry of the crystal and generically the crossing bands can hybridize and gap out the NL. This is demonstrated in Figure \ref{fig:SOC_xax_zoom}. From the orbital projection onto $d_{xy/x^2-y^2}$ and $d_{xz/yz}$ states one can clearly see the hybridization and gap between the valence band and one of the two near-degenerate conduction bands (the second conduction band passes through the gap). However, the gap is small (order of $10$ meV), and we expect qualitative differences from the collinear and [001] oriented cases to be negligible. Finally, we note that the magnetic anisotropy energy $E_{[001]}-E_{[100]}$ is relatively modest ([001] oriented spin is lower in energy than [100] by $18$ meV/f.u), implying that it should be feasible to switch between a robust and gapped nodal line within the nonpolar $P6_3/mmc$ space group by varying the direction of an external magnetic field. 

\subsection{Surface States}
\begin{figure*}
\subfigure[]{\label{fig:Vsurf}}\includegraphics{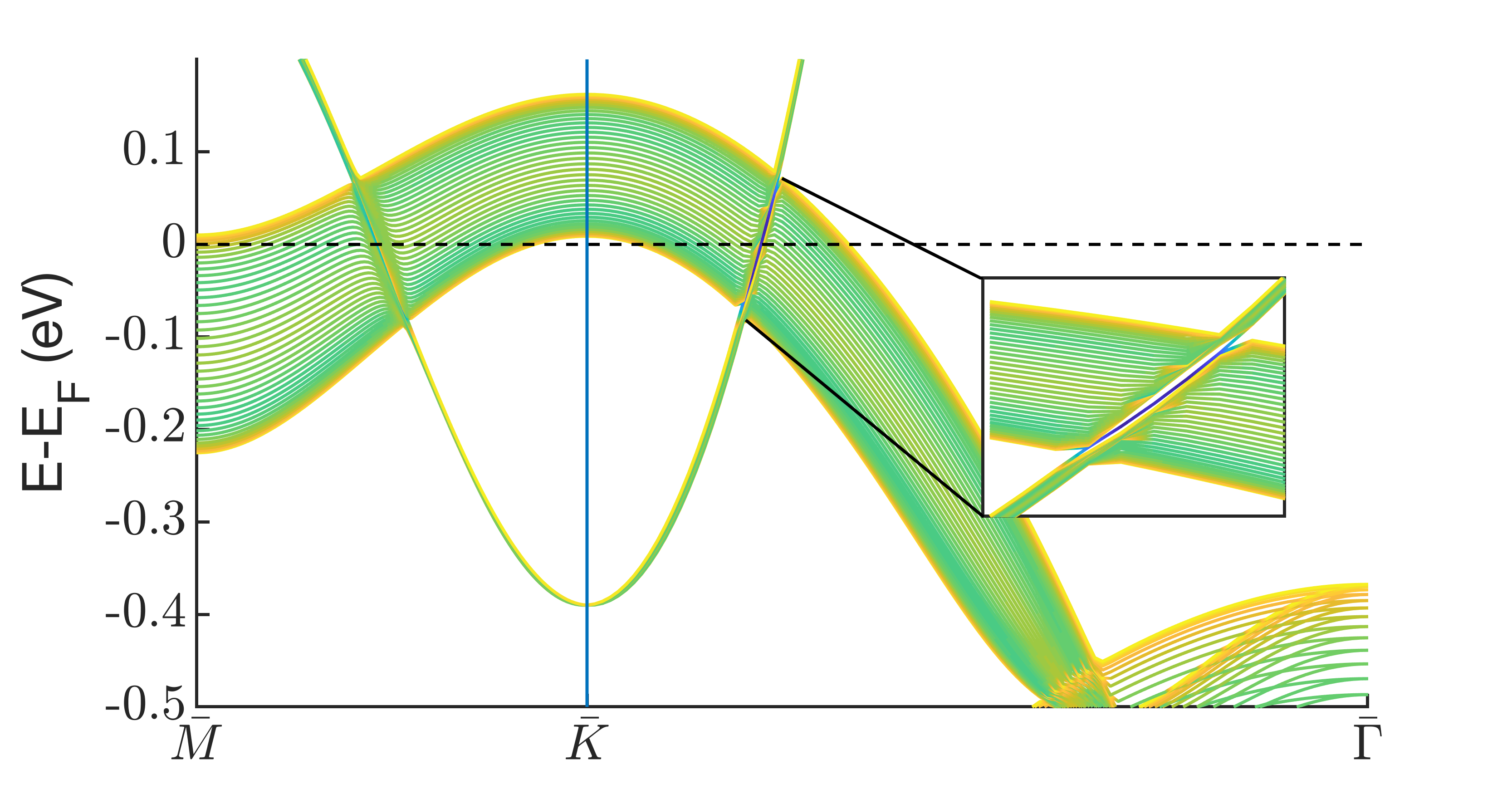}
\subfigure[]{\label{fig:Crsurf}}\includegraphics{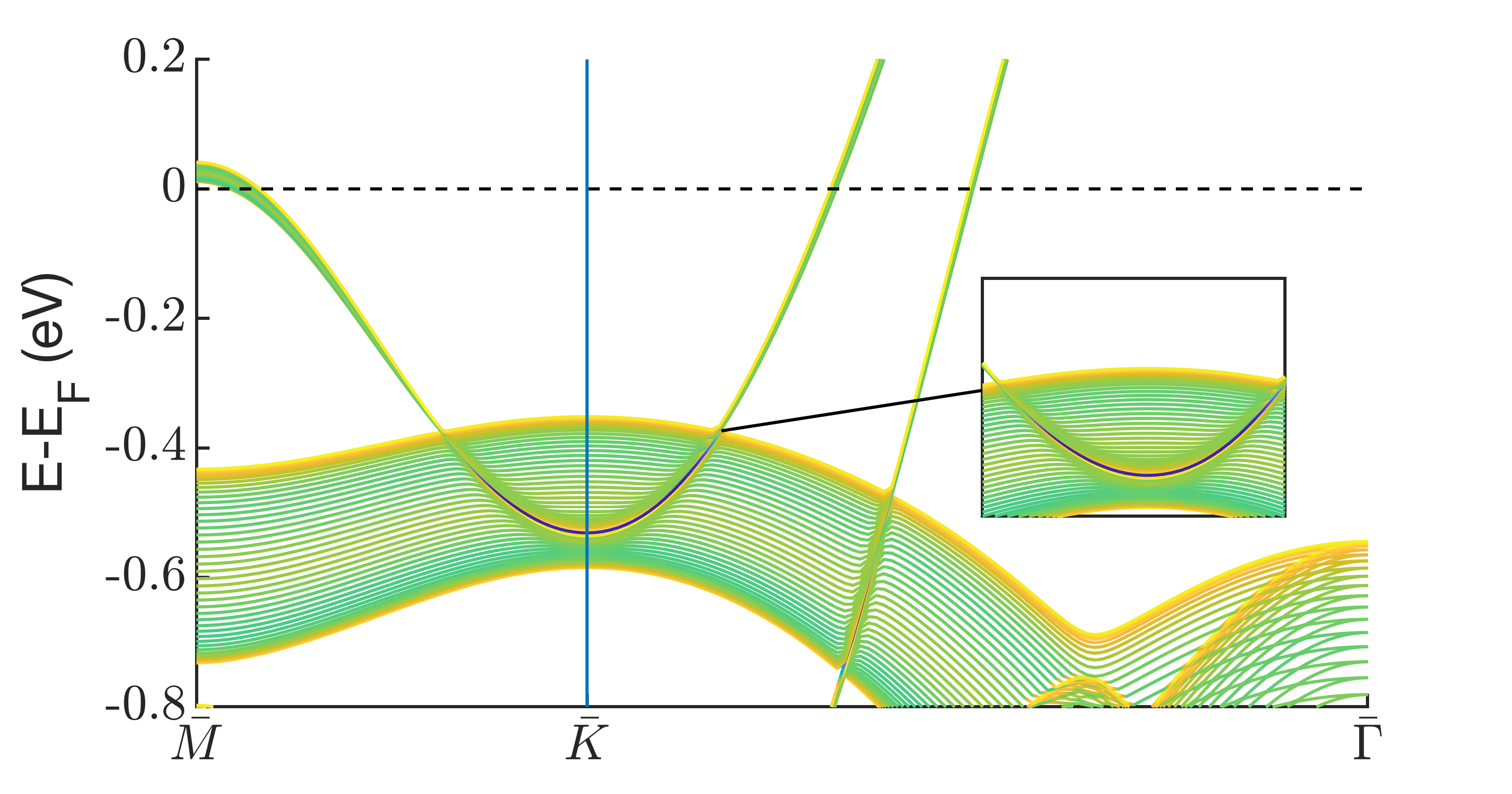}
\subfigure[]{\label{fig:surf_project}}\includegraphics[width=\columnwidth]{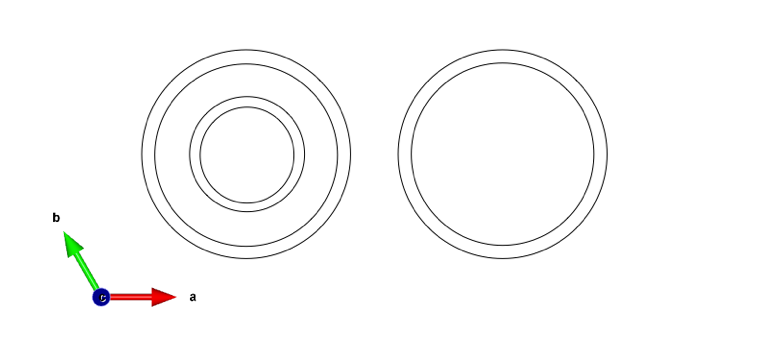}
\caption{\label{fig: surfstates} Projected band structures for the [001] surface in (a) $\mathrm{YVO_3}$ and (b) $\mathrm{YCrO_3}$ from slab geometries (see text). Color is proportional to weight of projection onto the outermost layers of the slab, with blue being highest weight. The discernible surface state is magnified in the inset for both cases. (c) Cartoon schematics of the multiple NL projections onto the [001] surface for $\mathrm{YVO_3}$ (left) and $\mathrm{YCrO_3}$ (right).}
\end{figure*}
A hallmark feature of topological NLs is their two-dimensional ``drumhead" surface states\cite{Chan2016}. These surface states must terminate at the surface projection of the nodal line and they may lie either outside or inside the area subtended by the NL. Using our maximally localized Wannier functions (MLWFs) derived from $\mathrm{V}$/$\mathrm{Cr}$ $\mathrm{d}$ states and $\mathrm{O}$ $\mathrm{p}$ states, we construct a slab model with 20 unit cells in the [001] direction. The 2D projected band structures on the $[001]$ surface for $\mathrm{YVO_3}$ and $\mathrm{YCrO_3}$ are shown in Figures \ref{fig:Vsurf} and \ref{fig:Crsurf}, respectively. At the the $K$ point, the top valence band of $\mathrm{YCrO_3}$ and top two valence bands of $\mathrm{YVO_3}$ invert with the bottom two conduction bands which are very nearly degenerate for both compounds. Thus, one (two) pair(s) of NLs are actually projected onto on the [001] surface in the case of $\mathrm{YCrO_3}$ and $\mathrm{YVO_3}$, respectively (see cartoon in Figure \ref{fig:surf_project}). While in principle there is a single surface drumhead state associated with each bulk NL\cite{Bian2016}, the projected bulk from the multiple NLs interferes with the surface states, making detection difficult. However, by projecting the tight-binding wave functions onto the outermost cells in the slab we can make out a single surface state (dark blue) which has not hybridized with bulk. It is sandwiched between the pair of NLs in $\mathrm{YCrO_3}$, whereas in $\mathrm{YVO_3}$ it is visible only in the region between the two NL pairs.

\subsection{Ferroelectric Instabilities of the $P6_3cm$ Structure}
\begin{figure}
\includegraphics[width=\columnwidth]{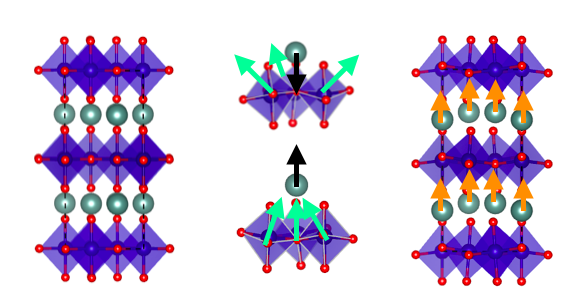}
\caption{Depiction of the nonpolar-to-polar structural transition in the hexagonal manganites. Left panel: Nonpolar centrosymmetric space group $P6_3/mmc$ (primitive cell tripled for easier comparison to polar phase). Middle panel: Action of the unstable $q=(1/3,1/3,0)$ $K_3$ phonon on the $\mathrm{XO_5}$ trigonal bipyramids. Outward trimerization pulls the $Y$ ions downwards (top), whereas inward trimerization forces the $Y$ ions upwards (bottom). Right panel: Polar $P6_3cm$ space group. The $K_3$ phonon couples to a secondary order parameter, the zone-centered $\Gamma_2^-$ mode (upward arrows), which further shifts the $Y$ ions in the $\hat{z}$ direction and causes net polarization in the unit cell.}
\label{fig:NP_to_P}
\end{figure}

\begin{figure*}
\subfigure[]{\label{fig:V}}\includegraphics{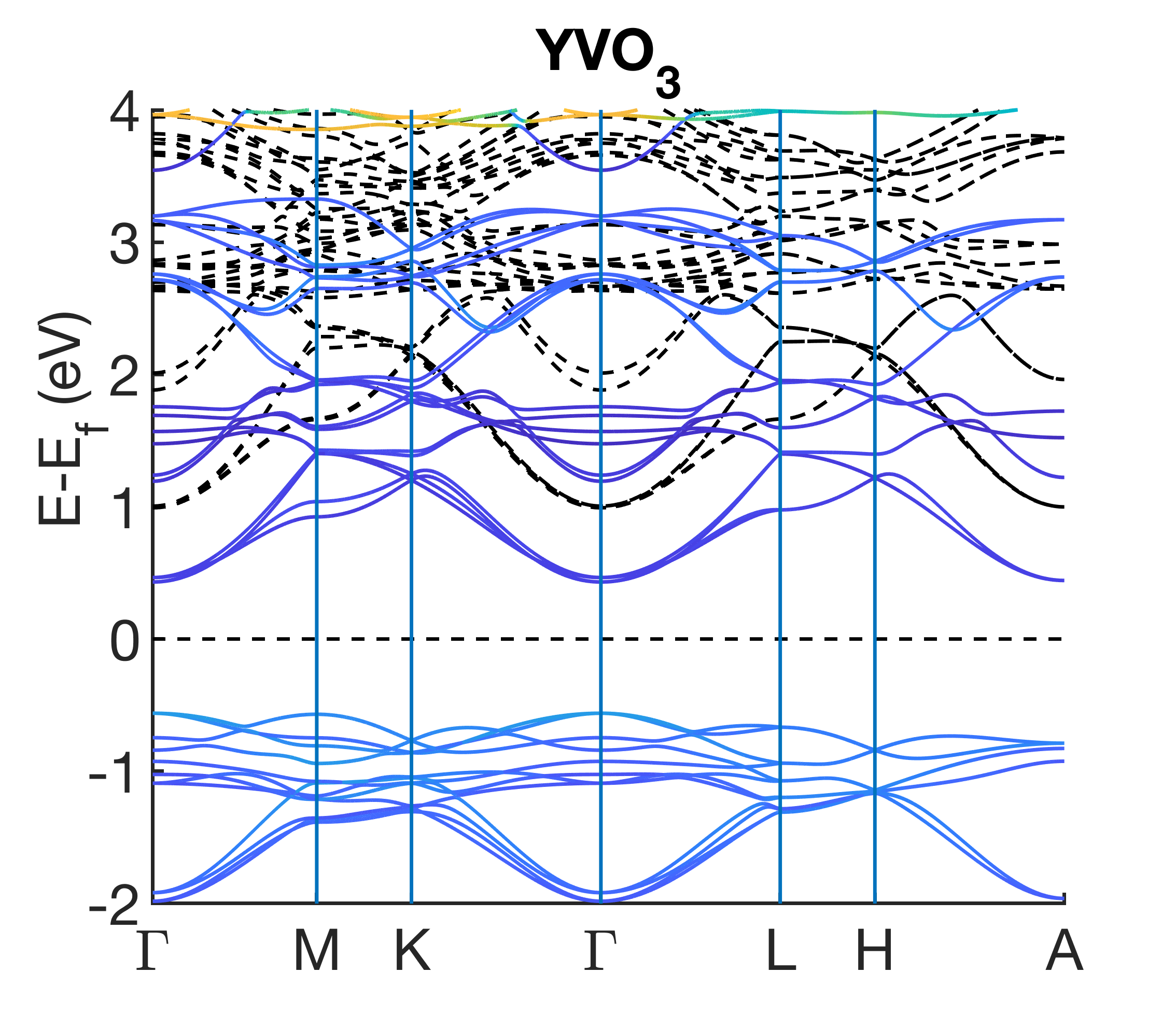}
\subfigure[]{\label{fig:Cr}}\includegraphics{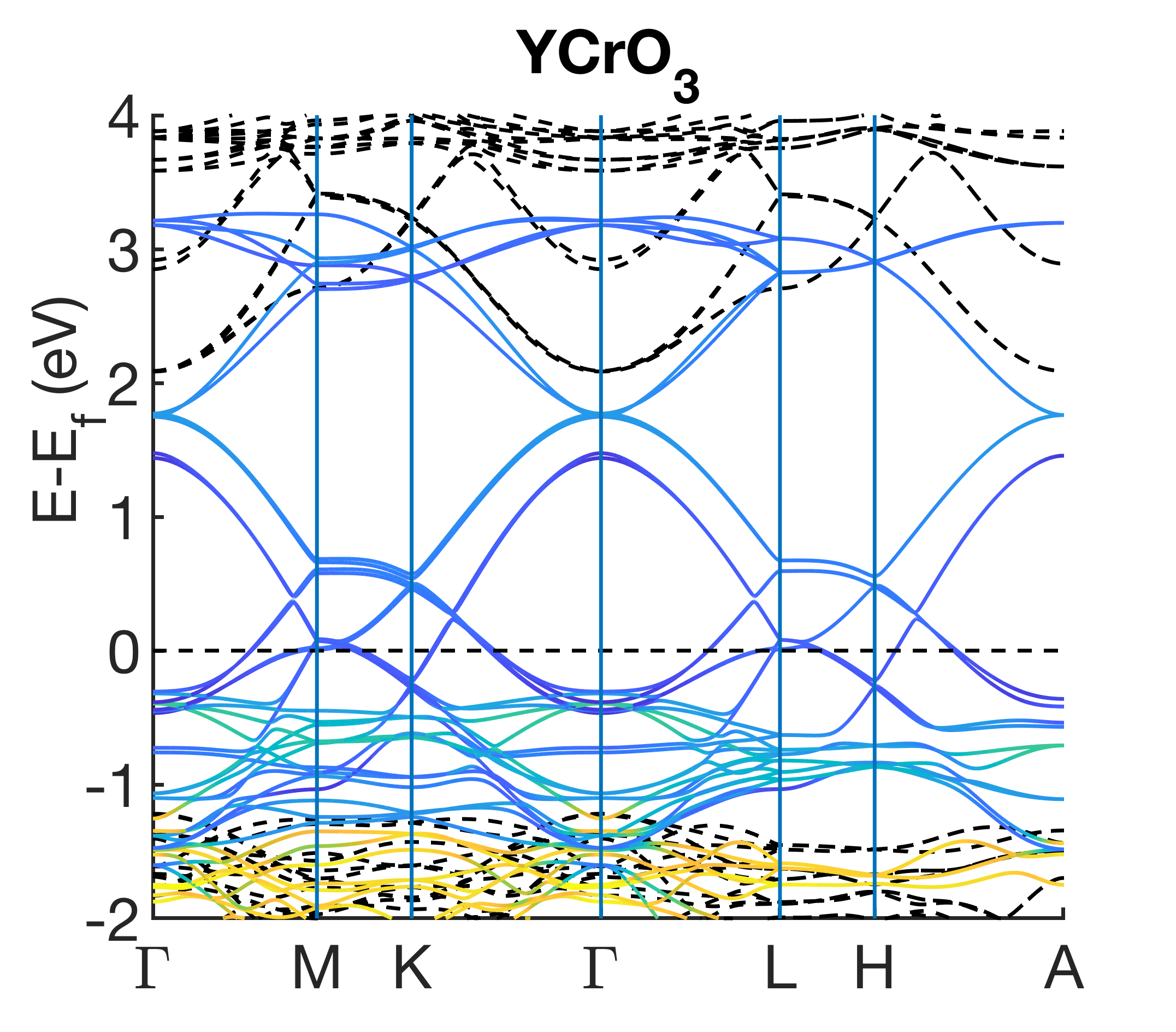}
\subfigure[]{\label{fig:Mn}}\includegraphics{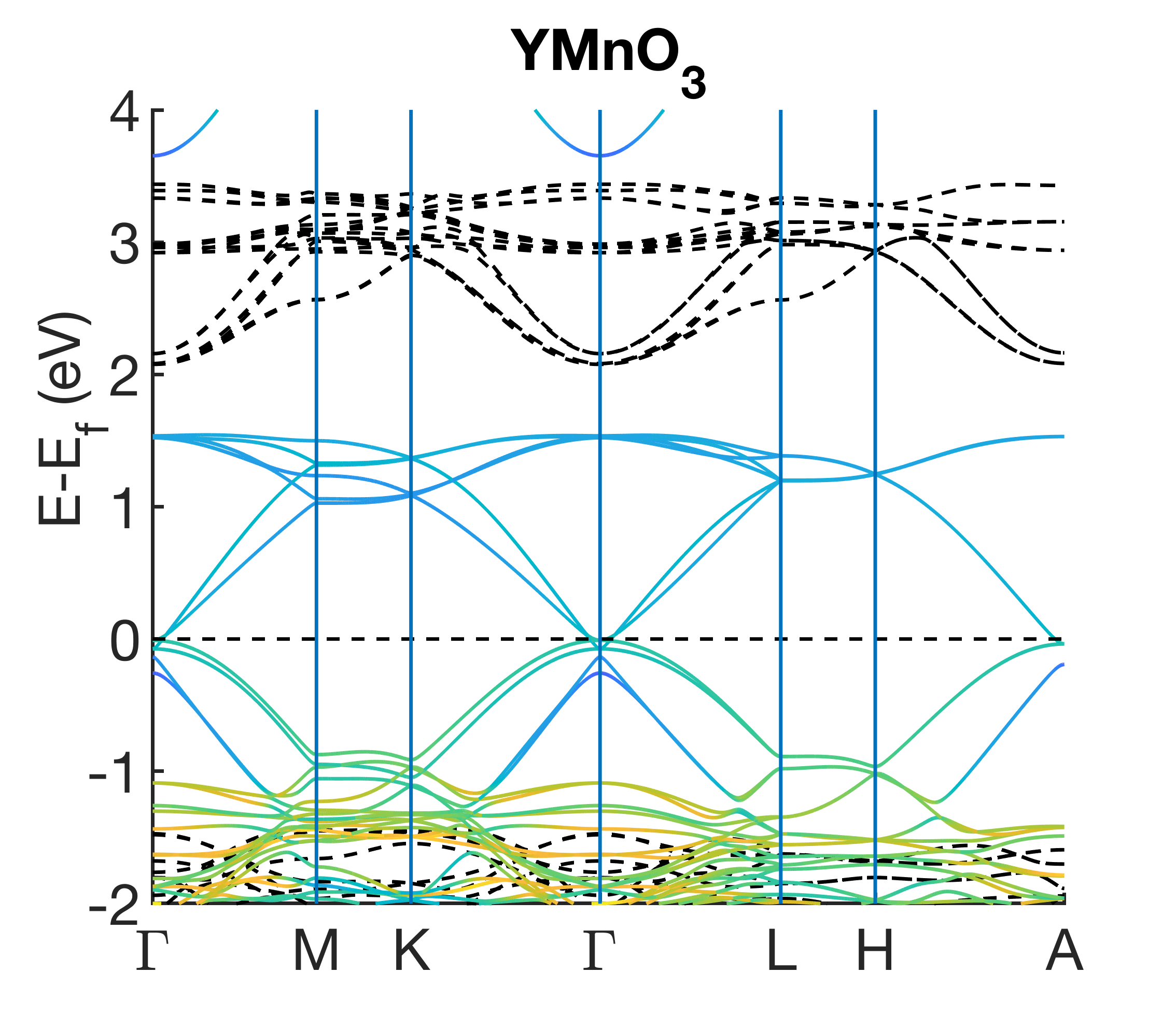}
\subfigure[]{\label{fig:Fe}}\includegraphics{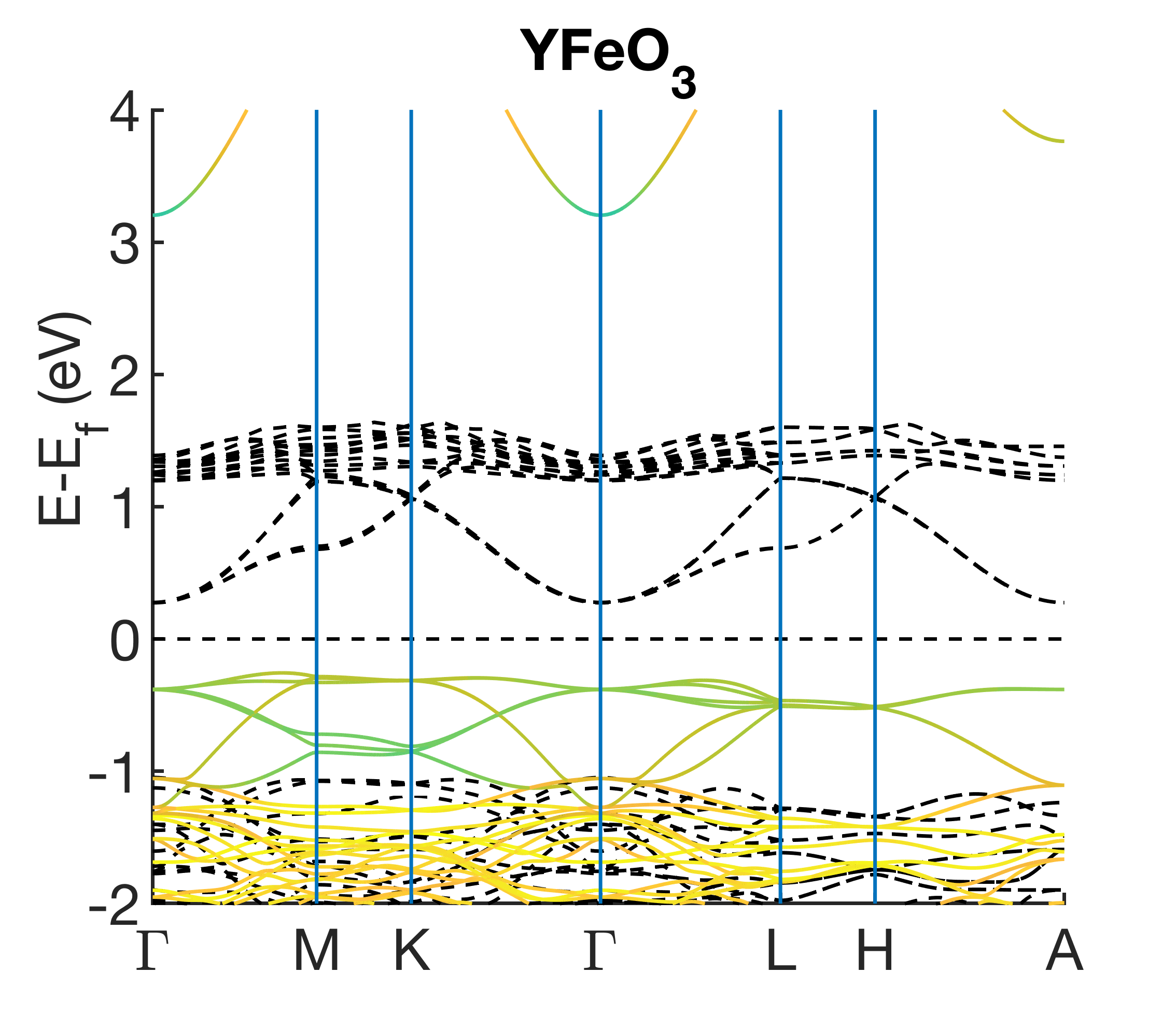}
\subfigure[]{\label{fig:Co}}\includegraphics{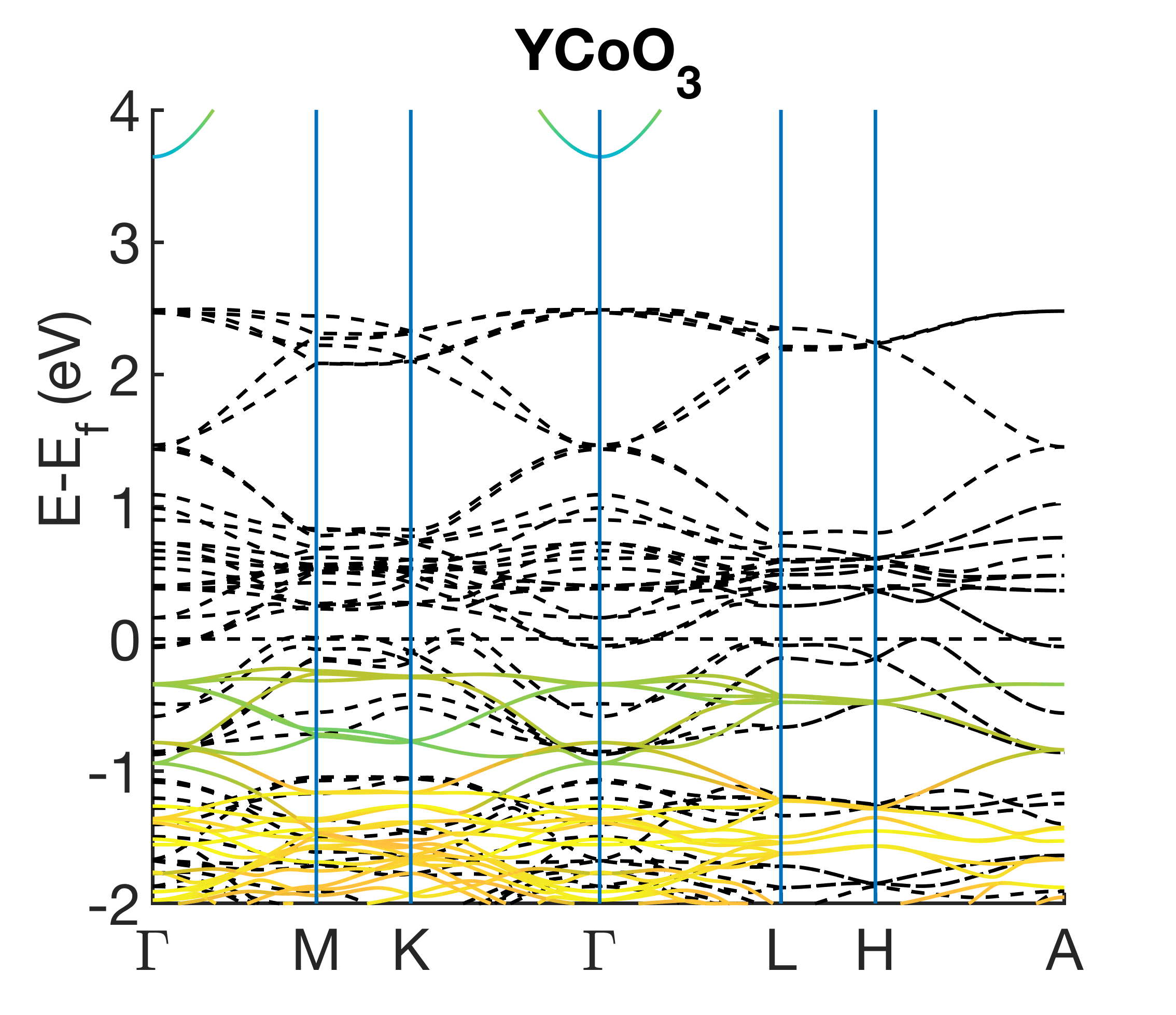}
\subfigure{\label{fig:bar}}\includegraphics{colbar.png}
\caption{\label{fig: polorbs} DFT-GGA+U band structures for the FM $\mathrm{YXO_3}$ compounds (X=V-Co) in the polar $P6_3cm$ space group . For comparison to Figure \ref{fig: orbplots}, the spin-up bands are again projected onto the relevant atomic orbitals and the spin-down bands are plotted in dashed black. Panels (a)-(e) correspond to $\mathrm{YVO_3}$, $\mathrm{YCrO_3}$, $\mathrm{YMnO_3}$, $\mathrm{YFeO_3}$, and $\mathrm{YCoO_3}$, respectively.}
\end{figure*}

The topological NLs near the Fermi level occur in the high-symmetry $P6_3/mmc$ space group due to the combination of band inversion at $K$ and the $\mathcal{R}_z$ mirror symmetry. However, as mentioned previously the hexagonal manganites $\mathrm{RMnO_3}$ are known to undergo a ferroelectric (FE) transition to the nonpolar $P6_3cm$ space group as the temperature is lowered\cite{Lonkai2004}. Here we verify that the $\mathrm{YXO_3}$ ($\mathrm{X}=\mathrm{V}$-$\mathrm{Co}$) compounds in their metastable hexagonal structure also have a lower-energy $P6_3cm $ phase connected to the $P6_3/mmc$ topological semimetal phase through unstable phonon modes.\\
\indent We first compute the energy per formula unit of the FM $P6_3cm$ polar structures and compare with our previously calculated energies for the FM $P6_3/mmc$ nonpolar structures. The GGA+U $\Delta E=E_{polar}-E_{nonpolar}$ is given in Table \ref{tab:mode_and_E}. For all five compounds the polar phase is lower in energy. Next, we decompose the atomic displacements involved in the distortion from the nonpolar to the polar phase into symmetry-adapted phonon modes using the AMPLIMODES software\cite{Orobengoa2009, Perez-Mato2010} provided by the Bilbao Crystallographic Server. The primary structural order parameter responsible for the $P6_3/mmc\rightarrow P6_3cm$ transition in hexagonal manganites is the unit-cell tripling $K_3$ phonon mode at $q=(\frac{1}{3},\frac{1}{3},0)$\cite{Lonkai2004,Fennie2005}. As temperature is lowered this phonon can condense, leading to trimerizing tilts of the $\mathrm{XO_5}$ trigonal bipyramids and a subsequent shifting either up or down of the $\mathrm{Y}$ ions, as shown in the middle panel of Figure \ref{fig:NP_to_P}. At this point the $P6_3/mmc\rightarrow P6_3cm$ transition has already occurred, but there is no \emph{net} polarization in the unit cell. The spontaneous polarization is caused by the coupling to $K_3$ of the zone-centered $\Gamma_2^-$ mode at $q=(0,0,0)$. $\Gamma_2^-$ causes an additional uniform shift of the $\mathrm{Y}$ ions in the $\hat{z}$ direction, resulting in non-zero polarization\cite{Griffin2014} (right panel of Figure \ref{fig:NP_to_P}). Based on the relative amplitudes of the modes (given in \AA) in the $P6_3cm$ structures relative to the parent $P6_3/mmc$ structures in Table \ref{tab:mode_and_E}, we can conclude that the FE transitions in the $\mathrm{YXO_3}$ compounds of interest also exhibit the $K_3$ mode as their primary order parameter, with the distortion caused by the $\Gamma_2^-$ mode significantly smaller. Moreover, the $K_3$ distortion amplitudes are all modest in magnitude, roughly $1$ \AA, implying that the FE transition is realistic for these systems.\\
\indent Finally, we briefly examine the band structures for the fully relaxed $P6_3cm$ compounds in Figure \ref{fig: polorbs}. In addition to the loss of inversion symmetry in the nonpolar-to-polar transition, the $\mathcal{R}_z$ symmetry protecting the NLs in the $P6_3/mmc$ space group is no longer a symmetry for $P6_3cm$. Thus, the topological NLs in $P6_3/mmc$ $\mathrm{YVO_3}$ and $\mathrm{YCrO_3}$ are necessarily absent. According to our band structure calculations, $\mathrm{YCrO_3}$ becomes a trivial metal. $\mathrm{YVO_3}$ on the other hand develops a $~1$ eV direct gap in Figure \ref{fig:V}. Since this would allow a tuning between a topological semimetal state and a trivial insulator by changing temperature, $\mathrm{YVO_3}$ seems to be the most promising of the $\mathrm{YXO_3}$ candidates for future studies.

 \begin{table}
 \caption{\label{tab:mode_and_E} $\Delta E=E_{polar}-E_{nonpolar}$ and amplitudes of $K_3$ and $\Gamma_2^-$ modes of the polar $P6_3cm$ structure with respect to $P6_3/mmc$ parent structure. Note that all calculations here are with FM ordering.}
 \begin{ruledtabular}
 \begin{tabular}{| c | c | c | c | c | c |}
 \hline
 & $\mathrm{YVO_3}$ & $\mathrm{YCrO_3}$ & $\mathrm{YMnO_3}$ & $\mathrm{YFeO_3}$ & $\mathrm{YCoO_3}$ \\ \hline
$\Delta E$ (eV) & $-0.354$ & $-0.075$ & $-0.100$ & $-0.103$ & $-0.358$\\ \hline
$K_3$ (\AA) & $1.056$ & $0.958$ & $0.971$ & $1.030$ & $1.026$\\ \hline
$\Gamma_2^-$ (\AA) & $0.402$ & $0.158$ & $0.183$ & $0.199$ & $0.218$\\ 
\hline
\end{tabular}
\end{ruledtabular}
\end{table}

\section{Conclusion}
In summary we have performed extensive first-principles calculations on five $\mathrm{YXO_3}$ (X=V-Co) compounds isostructural to the hexagonal manganite $\mathrm{YMnO_3}$. We find that with FM ordering the nonpolar $P6_3/mmc$ phase hosts topologically nontrivial nodal lines near the Fermi level for $\mathrm{YVO_3}$ and $\mathrm{YCrO_3}$. The NLs are formed by a band inversion and protected by a mirror plane symmetry. We show that the $\mathrm{YXO_3}$ compounds are also ferroelectric, undergoing a structural transition to polar $P6_3cm$ upon lowering of temperature. Finally, $\mathrm{YVO_3}$ becomes insulating in the polar phase, suggesting the possibility of switching from a TSM to an insulating state concomitantly with the FE transition. In realizing these structures, FM magnetic order must be stabilized in the nonpolar space group; in principle this could be done via application of a magnetic, electric, or even strain field\cite{Fennie2006}. Although all compounds except $\mathrm{YMnO_3}$ naturally crystallize in an orthorhombic structure, rather than the hexagonal phase studied here\cite{Communication2007,Maiti2010,Michel2004, Tsvetkov2004}, it is possible to synthesize a metastable structure by epitaxial growth on a hexagonal substrate. In fact, this has already been done successfully for the case of $\mathrm{YFeO_3}$\cite{Ahn2013}. Thus, our studies provide motivation for future experimental work stabilizing the hexagonal FM phases, thereby providing a new opportunity for examining the interplay of multiferroicity and topology.

\begin{acknowledgments}
This work is supported by the Center for Novel Pathways to Quantum Coherence in Materials, an Energy Frontier Research Center funded by the US Department of Energy, Director, Office of Science, Office of Basic Energy Sciences under Contract No. DE-AC02-05CH11231. Computational resources provided by the Molecular Foundry through the US Department of Energy, Office of Basic Energy Sciences, and the National Energy Research Scientific Computing Center (NERSC), under the same contract number. S. F. W. was supported under the National Defense Science and Engineering Graduate Fellowship (NDSEG). Calculations were performed on the Lawrencium cluster, operated by Lawrence Berkeley National Laboratory, and on the National Energy Research Scientific Computing Center (NERSC). 
\end{acknowledgments}

\FloatBarrier
\bibliography{Top_Hex_Mang_paper.bib}

\end{document}